\documentclass{article}
\usepackage{fancyhdr}
\usepackage{arxiv}
\usepackage[utf8]{inputenc}
\usepackage[T1]{fontenc}
\usepackage{hyperref}
\usepackage{url}
\usepackage{booktabs}
\usepackage{amsfonts}
\usepackage{nicefrac}
\usepackage{microtype}
\usepackage{multirow}
\usepackage{tcolorbox}
\tcbuselibrary{breakable}
\usepackage{algorithm}
\usepackage{algpseudocode}
\usepackage{amsmath}
\usepackage{graphicx}
\usepackage[natbibapa]{apacite}
\usepackage{doi}
\usepackage{tikz}
\usepackage{titlesec}
\usetikzlibrary{positioning,arrows.meta,fit,shapes.geometric,calc,backgrounds}

\title{Evaluating open-weight LLMs for scoring L2 essays: Accuracy, prompt transfer, and deployment of AiAWE}
\date{}
\author{John Maurice Gayed\\
\normalfont Global Education Center, Waseda University, Tokyo, Japan\\
\normalfont ORCID: \url{https://orcid.org/0000-0002-7028-3291}\\}

\hypersetup{
pdftitle={Evaluating open-weight LLMs for scoring L2 essays: Accuracy, prompt transfer, and deployment of AiAWE},
pdfsubject={cs.CL, cs.AI},
pdfauthor={John Maurice Gayed},
pdfkeywords={Automated Writing Evaluation, AI in Education, Open-Weight LLMs, LoRA, Essay Scoring},
}

\begin{document}

\maketitle

\begin{abstract}
This study presents AiAWE, an open-source automated writing evaluation system that scores argumentative essays with a LoRA-adapted large language model deployed on local hardware. Using 480 TOEFL Independent Writing essays from a restricted-access Educational Testing Service (ETS) dataset, we fine-tuned Gemma-3-27B and LLaMA-3.3-70B under identical LoRA configurations on 120 essays from one prompt, with all hyperparameters fixed in advance, and evaluated them on the remaining 360 essays. Gemma achieved a root mean square error (RMSE) of 0.474 and a quadratic weighted kappa (QWK) of 0.828 on the half-point grid against the ETS scores. In paired comparisons, it significantly outperformed a fine-tuned GPT-3.5 baseline from prior work on every metric tested, showing that a locally deployable open-weight model can achieve competitive scoring performance in this setting. It also had significantly lower error than the larger LLaMA configuration, although, because the two models differ in family as well as size, this comparison does not isolate the effect of scale. Neither open-weight model showed a statistically detectable increase in error on the 240 essays written to a second, unseen prompt from the same task family. On a single consumer-grade GPU, the scoring model returned a complete response in a median of 14.2~s, demonstrating technical feasibility under the tested conditions. Only the scoring component was evaluated, not the generated feedback. The system is publicly accessible at \url{https://app.awade.gec.waseda.ac.jp}, and its code, LoRA adapters, and training configurations are openly released.
\end{abstract}


\section{Introduction}
\label{sec:Intro}

Automated writing evaluation (AWE) has become one of the main applications of large language models (LLMs) in education. Writing assessment is labor-intensive, subject to rater fatigue and inconsistency, and increasingly difficult to scale as assessment and feedback volumes grow \citep{emirtekin2025large, ramesh2022automated}. LLMs, with their strong instruction-following and text-understanding capabilities, offer a plausible path toward systems that can score essays at human-comparable reliability while producing the kind of rubric-aligned qualitative feedback that formative assessment requires. Recent work has shown that fine-tuning proprietary LLMs such as OpenAI's GPT-3.5 Turbo and GPT-4 on even modest quantities of human-scored data produces substantial improvements \citep{wang2024effectiveness, liu2025comparing, latif2024finetuning}, and that these fine-tuned models can approach or match human inter-rater agreement on standard benchmarks \citep{wang2024effectiveness, liu2025enhancing}.

However, much of the leading empirical work in this area relies on proprietary APIs, most often OpenAI's GPT family. This reliance raises three practical concerns that have been flagged repeatedly in recent literature. First, sending student essays to commercial APIs moves sensitive educational data outside institutional control, creating compliance risks for institutions operating under GDPR, FERPA, or equivalent regimes \citep{ormerod2024automated}. Second, closed-source model weights prevent the kind of auditability that high-stakes assessment requires. Researchers cannot inspect what the model has learned, measure its biases directly, or verify that its behavior is stable over time. Third, reproducibility is undermined when model versions change silently (a common occurrence with commercial systems) and fine-tuning procedures are not fully disclosed \citep{ormerod2024automated}. The alternative, using open-weight LLMs, has lagged behind in the AWE literature, with fewer published studies measuring the effectiveness of open models on scoring tasks.

This gap motivates the present study. We build on the methodology of \citet{wang2024effectiveness}, who fine-tuned GPT-3.5 Turbo on the ETS TOEFL Independent Writing dataset and established that task-specific fine-tuning substantially improves scoring accuracy over zero-shot prompting, achieving a root mean square error (RMSE) of 0.57 and a quadratic weighted kappa (QWK) of 0.78. This project's goal is to determine whether the same benchmark can be met or exceeded using open-weight models and a fully open-source deployment pipeline. If it can, this would show that, for the task examined here, a locally deployable model is a viable option for institutions that cannot send student writing to commercial APIs.

To this end, we fine-tune two instruction-tuned open-weight models, Google's Gemma-3-27B-it and Meta's LLaMA-3.3-70B-Instruct, using Low-Rank Adaptation (LoRA) \citep{hu2022lora} on the same 480-essay ETS dataset employed in \citet{wang2024effectiveness}. Under a 120-train/360-test split with identical LoRA hyperparameters for both models, we evaluate scoring accuracy against the ETS human benchmark scores and against each other. The fine-tuned models are deployed using \texttt{llama.cpp} on a local workstation equipped with dual consumer-grade RTX 3090 GPUs, demonstrating that the inference pipeline, which is what an institution needs to run the system, is operable on hardware accessible to many universities and independent researchers. Fine-tuning itself was performed on a rented data-center GPU (Section~\ref{sec:training_infra}). The resulting system, AiAWE, is publicly accessible at \url{https://app.awade.gec.waseda.ac.jp} and serves students and educators with essay scoring, rubric-referenced qualitative feedback, batch processing, and custom-rubric configuration. The present study evaluates the scoring component of this system. The qualitative feedback it generates is described but not evaluated here (Section~\ref{sec:limitations}).

A factor that shapes several of our findings is the use of LoRA on an instruction-tuned checkpoint rather than full fine-tuning. Because LoRA freezes the base model's weights and learns only a small low-rank update at a moderate rank, the original model's general-purpose instruction-following and text-generation capabilities are largely preserved during adaptation \citep{hu2022lora, biderman2024lora}. As a result, a single LoRA-adapted model can produce both a numeric score and rubric-referenced qualitative feedback, the latter drawing on the preserved generative abilities of the base model. Only the first of these is evaluated in this study. Separately, we keep the learned adapters unmerged from the base weights at inference time. This is a deployment convenience rather than a determinant of the dual-purpose behavior, because the forward pass is the same whether or not the adapter is merged. It does, however, simplify distribution (a small adapter file rather than a full checkpoint), allow multiple task-specific adapters to be served on one frozen base, and avoid merging the adapter into the base and then re-quantizing it for our \texttt{llama.cpp} pipeline. Together, these choices allow researchers to adapt a large LLM to a specific downstream task with modest resources.

This study addresses four research questions:

\begin{description}
\item[RQ1] Can a locally deployable open-weight instruction-tuned LLM, adapted with LoRA on a small rubric-graded dataset, achieve scoring accuracy comparable to or better than a fine-tuned GPT-3.5 baseline on the same essays?

\item[RQ2] Under identical LoRA hyperparameters and training data, how does the scoring accuracy of LoRA-adapted LLaMA-3.3-70B compare with that of LoRA-adapted Gemma-3-27B?

\item[RQ3] Does scoring accuracy generalize to essays written in response to an essay prompt not seen during fine-tuning?

\item[RQ4] Is it technically feasible to run the fine-tuned scoring model on consumer-grade hardware, in terms of measured inference latency and throughput?
\end{description}

In brief, for RQ1, the Gemma-based system improved significantly on the GPT-3.5 baseline on every metric tested. For RQ2, the Gemma-3-27B configuration achieved significantly lower scoring error than the LLaMA-3.3-70B configuration, and a descriptively higher QWK, although the two did not differ significantly in exact or within-$\pm$0.5 agreement. Because the two models differ in family as well as in size, this is a comparison of two specific configurations rather than a test of model scale. For RQ3, neither open-weight model showed a statistically detectable increase in error on essays written to the unseen prompt. For RQ4, the scoring model emits its first token after 1.28~s and completes a full response in a median of 14.2~s on a single consumer GPU, which makes streamed interactive use and offline batch scoring technically feasible under the tested conditions.

\section{Related Work}
\label{sec:Related}

This section reviews three interconnected research threads that situate the present study: (1)~the evolution of automated essay scoring from feature-engineered and neural approaches to large language model (LLM)-based methods; (2)~parameter-efficient fine-tuning with Low-Rank Adaptation (LoRA) and the trade-offs that govern its effectiveness; and (3)~quantized and open-source deployment strategies that enable LLM inference on consumer-grade hardware.

\subsection{LLM-Based Automated Writing Evaluation}
\label{sec:llm_aes}
The application of LLMs to automated essay scoring (AES) and automated writing evaluation (AWE) has expanded rapidly since 2023, driven by the instruction-following and zero-shot capabilities of models in the GPT and Gemini families. Early investigations by \citet{mizumoto2023exploring} applied GPT-3 (\texttt{text-davinci-003}) to the 12,100 essays of the TOEFL11 corpus via zero-shot prompting. Because that corpus releases collapsed proficiency bands rather than numeric scores, the model was prompted with the IELTS Task 2 band descriptors and its 0--9 output was then collapsed to three levels for comparison. On that basis it matched the corpus proficiency label in 54.33\% of cases, which the authors characterize as low. The same study also asked whether established lexico-grammatical complexity indices could improve on the model's scores taken alone, and found that incorporating them did. Although the zero-shot results were a useful proof of concept, scoring programs that need higher reliability thresholds motivate subsequent work on fine-tuning and prompt engineering.

A foundational study for the present project is \citet{wang2024effectiveness}, who fine-tuned GPT-3.5 (turbo-1106) on the ETS TOEFL Independent Writing dataset, the same 480-essay corpus used in this study. The fine-tuned model achieved an RMSE of 0.57 and a QWK of 0.78, substantially outperforming both zero-shot GPT-3.5 and the more capable GPT-4 under identical conditions. Those results established two key findings: (a)~task-specific fine-tuning on rubric-aligned data yields substantial gains over prompting alone, and (b)~fine-tuned models do not require a large variety of essay prompts to generalize effectively. The first of these has since been corroborated at larger scale by \citet{liu2025comparing}, who compared six GPT-based approaches, including two fine-tuned models, two GPT-4 configurations and two GPT-3.5 configurations, and found that both fine-tuned models outperformed every non-fine-tuned variant. 

The linguistic-feature strand of \citet{mizumoto2023exploring} has been taken further by \citet{liu2025enhancing}, working on the same TOEFL11 corpus. They fine-tuned a GPT model to rate essays from four L1 groups and compared three approaches: GPT ratings only, 22 linguistic complexity indices only, and the two combined. The fine-tuned model reached 78.3\% exact agreement with human raters overall. GPT ratings outperformed the complexity indices, and combining the two yielded only marginal further improvement. Three additional results concern fairness. Reliability and accuracy were lower for low-scoring essays, a minority class in the training data. The model showed a general tendency to overrate. In addition, its performance varied across the four L1 groups, with the lowest reliability and accuracy observed for L1 German writers, which the authors frame explicitly as a validity concern for examinees from diverse language backgrounds and as grounds for addressing L1-related bias in AES. 

Beyond the GPT family, \citet{atkinson2025hybrid} proposed a hybrid AES architecture that combines LLM-derived features with discourse-level and structural representations, demonstrating that multi-feature integration can outperform pure neural approaches on standard essay datasets. \citet{liew2024automated} evaluated GPT-4, GPT-3.5, PaLM, and LLaMA-2 on both scoring and feedback generation tasks, reporting a QWK of 0.68 and qualitative feedback agreement scores of 4.9 out of 5.0 with human examiners. Their work highlighted the dual utility of LLMs for simultaneous scoring and feedback, a design principle that informs the AiAWE architecture. The reliability of LLM-based scoring has been examined across multiple dimensions. \citet{tang2024harnessing} investigated the effects of prompt engineering, temperature settings, and multi-dimensional rubric criteria on GPT-4 and GPT-3.5 scoring performance, finding that lower temperatures produced more consistent scores and that GPT-4 achieved the highest reliability on the Ideas and Organization dimensions (QWK $= 0.551$ and $0.584$, respectively). \citet{latif2024finetuning} demonstrated that fine-tuned GPT-3.5 outperformed fine-tuned BERT by an average of 9.1\% in scoring accuracy across six science assessment tasks, further validating the advantage of generative LLMs over discriminative models for constructed-response evaluation.

Two recent systematic reviews consolidate these trends. \citet{emirtekin2025large} analyzed 49 peer-reviewed studies on LLM-powered automated assessment published between 2018 and 2024. The review found that while models such as GPT-4 can achieve very high agreement with human raters (e.g., QWK up to 0.99 in some configurations), performance varies considerably across tasks and scoring contexts, with some studies reporting ICCs as low as 0.45. \citet{ramesh2022automated} surveyed deep learning-based AES and feedback generation, tracing the evolution from BERT-based discriminative scorers to generative LLM approaches. The survey noted that despite strong scoring results, most generative models have not yet been deployed in live educational environments. A common thread across these studies is that zero-shot and few-shot LLM scoring, while impressive, does not yet match the reliability of fine-tuned models for rubric-aligned assessment. Fine-tuning on even modest quantities of human-scored data (see \citealp{wang2024effectiveness}) produces substantial and consistent improvements. These studies also indicate that what a model responds to, and how it behaves across writer groups, are relevant to its validity.

\subsection{Validity, Fairness, and the Evaluation of Automated Scores}
\label{sec:validity}

Accuracy statistics of the kind reported above are necessary but not sufficient evidence that an automated scoring system is fit for use. In the measurement literature, the relevant standard is a validity argument: an explicit, evidence-backed case that the interpretations and uses proposed for a set of scores are warranted \citep{kane2013validating}. On this account validity is a property of score interpretation and use rather than of the instrument itself, which means that a system adequate for one purpose (formative feedback on a draft) may be inadequate for another (placement or admissions) even with identical agreement statistics.

For automated scoring specifically, \citet{williamson2012framework} set out a reference framework, developed with ETS's e-rater as its worked example. It specifies the evidence an automated scoring program should assemble before operational use, including construct relevance and representation, agreement with human scores at defined thresholds, the behavior of the system across relevant subgroups, the consequences of the scores for test takers, and the procedures for ongoing monitoring once deployed. \citet{ramineni2013automated} elaborate the associated psychometric practices. \citet{ferrara2022validity} extend validity-argument thinking, in Kane's formulation, to scoring engines built on machine learning and observe that, because such models are often opaque, construct relevance must rest more on empirical evidence, including evidence of fairness across subgroups. \citet{xi2010automated} argues that automated scoring and feedback systems raise validation questions of their own and cannot simply inherit the validity argument established for human scoring.

Subgroup behavior is the dimension on which the LLM-based AES literature is currently thinnest. Differential performance by first-language background has been documented for pre-LLM automated scoring, and \citet{liu2025enhancing} report L1-linked variation in fine-tuned GPT scores. Since the writers whose essays such systems evaluate are, by construction, drawn from many language backgrounds, a scorer that is accurate in aggregate may still be systematically severe or lenient toward particular groups. Aggregate QWK is insensitive to exactly this shortcoming. We return to this point in Section~\ref{sec:limitations}, where we highlight the outcome of a companion fairness evaluation of the model described here. A second recurring finding concerns scoring severity. In a scoping review of 26 studies of GPT-based ratings of EFL writing, \citet{chen2026evaluating} reports that GPT tends to score more severely than human raters, particularly in language-related categories, alongside considerable heterogeneity in study design and reported outcomes. This pattern bears directly on the score-wise behavior we report in Section~\ref{sec:scorewise}.

\subsection{LoRA: Parameter-Efficient Fine-Tuning and Rank Sensitivity}
\label{sec:lora_lit}

Low-Rank Adaptation (LoRA), introduced by \citet{hu2022lora}, adapts a large model to a specific task by freezing the pre-trained weights and training only a small pair of low-rank ``adapter'' matrices injected into selected layers. At inference the adapter's contribution is added to the frozen base to produce task-specialized behavior. Two consequences matter here: training is inexpensive because only a small fraction of parameters receive gradient updates, and the base model's general capabilities remain intact. In our setting, one adapted model can in principle produce both a task-specific numeric score and general-purpose qualitative feedback. A generalized schematic of the mechanism is provided in Appendix~\ref{app:lora_fig}.

LoRA is not without trade-offs, and these have received substantial recent attention in the machine learning literature. A comprehensive empirical analysis by \citet{biderman2024lora} shows that relative to full fine-tuning (in which every parameter of the base model is updated), LoRA typically achieves lower accuracy on the target task but substantially better preservation of the base model's general capabilities. LoRA thus acts as a form of regularization. By constraining updates to a low-dimensional subspace, it prevents the fine-tuning process from overwriting representations that the base model learned during pre-training. This property suits the present use case, in which the model should acquire rubric-aligned scoring behavior without losing the ability to generate qualitative feedback.

A key variable in LoRA is the "rank" of the adapter, which controls how much capacity the adapter has to deviate from the base model. Lower ranks constrain the update to a smaller subspace (less capacity to fit the target task, but a tighter constraint on how far the model can move from its pretrained behavior), while higher ranks relax this constraint (more capacity to fit the task, but depending on how much the task's updates overlap with directions the base model relies on, a greater potential to disturb base capabilities). \citet{biderman2024lora} show that LoRA is sensitive to this rank setting and related hyperparameters, and that finding an appropriate value for a given task requires tuning against the desired output. In addition, \citet{rathore2025rank} conduct a rank sweep across reasoning and recall benchmarks and report that parameter-efficient adaptation can degrade capabilities outside the target domain at some rank settings. This literature shows that rank interacts with capability retention in ways that are not yet well characterized for generative instruction-tuned models.

Two design choices follow from this literature. First, rather than tuning aggressively, we use a moderate rank ($r = 64$) that the literature has found to be a reasonable balance point, with scaling factor $\alpha = 128$ following the widely adopted convention of $\alpha = 2r$ \citep{hu2022lora}. Second, we do not merge the LoRA adapters back into the base model weights after training. Instead, the adapters remain a separate file loaded alongside the frozen base at inference time, which simplifies distribution and allows multiple task-specific adapters to be served on one frozen base if needed. Finally, a related method is QLoRA \citep{dettmers2024qlora}, which extends LoRA by also quantizing the frozen base model to 4-bit precision during training. This reduces training memory further but introduces quantization noise into the gradient computation. We do not use QLoRA for training in this study. We fine-tune with the base model held in unquantized bf16 (16-bit) precision and apply quantization only at inference time. This is the standard practice recommended by recent work on open-source LLM deployment \citep{kurt2026quantization}.

\subsection{Quantized Inference and Open-Source Deployment}
\label{sec:deployment_lit}

The practical deployment of fine-tuned LLMs on consumer-grade hardware requires efficient inference frameworks and aggressive model compression. The \texttt{llama.cpp} project is a widely used open-source inference runtime for quantized LLMs, supporting both CPU and GPU backends with multiple quantization formats ranging from 1.5-bit to 8-bit precision. Its GGUF file format has become a widely used standard for distributing quantized model weights, and the framework supports tensor-level parallelism across multiple GPUs, a feature we exploit in this study to distribute the 70B-parameter LLaMA model across two RTX 3090 GPUs (48 GB of available VRAM). \citet{huang2024empirical} conducted a systematic evaluation of LLaMA-3 quantization across multiple bit-widths and methods, finding that 6-bit quantization preserves most of the model's capabilities while substantially reducing memory requirements. More aggressive 4-bit quantization introduces measurable but often acceptable degradation, particularly for tasks that do not require fine-grained numerical reasoning. \citet{kurt2026quantization} provided a unified evaluation of \texttt{llama.cpp} quantization schemes on Llama-3.1-8B-Instruct, confirming that Q6\_K and Q4\_K\_M represent effective trade-off points between compression and quality. These findings inform our choice of Q4\_K\_M for both Gemma and LLaMA at inference time (described in Section~\ref{sec:inference}), which we selected to enable an accuracy comparison between the two models at consumer-grade VRAM budgets.

The open-source pipeline adopted here, comprising open-weight models (Gemma, LLaMA), open tools (\texttt{llama.cpp}, LLaMA-Factory), and self-hosted inference, directly addresses the data-sovereignty, auditability, and cost concerns that \citet{ormerod2024automated} raise against closed-source educational scoring. Similar open-source approaches have been used successfully in other domains. \citet{gao2025enhancing} show that LoRA-adapted open-source LLMs can match or exceed proprietary models on clinical classification while preserving data sovereignty, and \citet{diez2024web} developed a web application for cost-effective fine-tuning of open-source LLMs in education.

\section{Methodology}
\label{sec:methodology}

This section describes the dataset, model selection, fine-tuning procedure, inference deployment, and evaluation metrics used to develop and assess the AiAWE essay scoring system.

\subsection{Dataset}
\label{sec:dataset}

The dataset consists of 480 argumentative essays written in response to two official TOEFL Independent Writing prompts, drawn from the TOEFL iBT Public Use Dataset provided by the Educational Testing Service (ETS). Access is restricted to ETS-approved researchers. The essays were written by 480 distinct test takers, so no writer appears more than once and writer independence across partitions is guaranteed by construction. The same 480 essays were used by \citet{wang2024effectiveness}, whose fine-tuned GPT-3.5 results serve as our comparison baseline. We note that this dataset is distinct from the TOEFL11 corpus \citep{blanchard2013toefl11}, which releases collapsed proficiency bands (low/medium/high) rather than numeric scores.

 TOEFL Independent Writing responses are scored on a 0--5 scale by trained human raters. To clarify the ETS rating procedure, we state it in full, following \citet{blanchard2013toefl11}. Two raters score each response. If the two scores are identical or adjacent (differing by no more than one point), they are averaged. If they differ by more than one point, a third rater scores the response. When the three scores form a consecutive sequence (no more than one point between neighboring scores), all three are averaged. When one score is an outlier, the two adjacent scores are averaged and the outlier discarded. In the single non-adjacent case (1, 3, 5) a fourth adjudicated score becomes final, and any response receiving a zero from any rater is routed to adjudication, where the adjudicated score is final. Averaging pairs of integers is what produces the 0.5 increments in the resulting scale.

Two properties of this criterion affect the interpretation of our results. First, it is a smoothed composite: averaging two or more ratings reduces rater-specific error variance, so the composite has lower error variance than any single human rating. Agreement between any model and this composite is therefore not the same quantity as agreement between two independent raters. Second, the dataset contains only the final composite. The individual rater scores are not distributed. We are consequently unable to report model agreement against rater 1 and rater 2 separately, or human--human agreement computed on the essays. We flag this as a limitation in Section~\ref{sec:limitations} rather than substituting an external human-agreement figure computed on a different scale and sample. Essays receiving a final score of zero (non-responsive or off-topic) are not present in the released subset, yielding a usable range of 1.0 to 5.0. The 480 essays are evenly split between the two prompts (240 each), and the score distribution is shown in Table~\ref{tab:score_distribution}.

\begin{table}[!h]
\centering
\caption{Score distribution across prompts.}
\label{tab:score_distribution}
\begin{tabular}{cccc}
\toprule
\textbf{Score} & \textbf{Prompt 1} & \textbf{Prompt 2} & \textbf{Total} \\
\midrule
1.0 & 1 & 2 & 3 \\
1.5 & 3 & 2 & 5 \\
2.0 & 18 & 20 & 38 \\
2.5 & 32 & 25 & 57 \\
3.0 & 65 & 55 & 120 \\
3.5 & 40 & 45 & 85 \\
4.0 & 32 & 38 & 70 \\
4.5 & 35 & 28 & 63 \\
5.0 & 14 & 25 & 39 \\
\midrule
\textbf{Total} & 240 & 240 & 480 \\
\bottomrule
\end{tabular}
\end{table}

Essays vary in length from 45 to 380 words (mean $\approx$ 214). After tokenization, input lengths ranged from 60 to 920 tokens, well within the 2048-token maximum used during training. Two experimental configurations were applied: (1)~$\mathcal{D}_{120}$, a 120-essay training set with 360 essays held out for evaluation, used to assess generalization from limited data; and (2)~$\mathcal{D}_{\text{full}}$, all 480 essays used for training the production model. No external or synthetic data were added.

\subsubsection{Data-splitting procedure}
\label{sec:split}

The fine-tuning data were selected in two steps. First, the data were stratified by score: we drew approximately 17 essays from each of the nine score levels, taking all available essays at levels where fewer than 17 exist. This deliberately over-samples the sparse tails relative to their population frequency, on the reasoning that fine-tuning benefits from coverage of the extremes. Second, all 120 training essays were drawn from prompt 1 only. This choice is inherited from \citet{wang2024effectiveness}, who trained separate GPT models on prompt 1, on prompt 2, and on both, and found that the prompt-1 model was the best performer: it significantly outperformed the two-prompt model and its scores (RMSE 0.57, QWK 0.78, 84.72\% agreement) are the GPT-3.5 baseline we compare against throughout. Training our open-weight models on the same prompt-1 subset therefore holds the training regime fixed against the strongest available baseline, so that any difference in Section~\ref{sec:results} is attributable to the model rather than to the essays or prompts it was trained on. Consequently, the held-out set comprises the 120 remaining prompt-1 essays and all 240 prompt-2 essays, and the models never observed a prompt-2 essay during training. Table~\ref{tab:split} gives the full composition.

\begin{table}[!h]
\centering
\caption{Composition of the $\mathcal{D}_{120}$ partition by score level and prompt. All training essays are drawn from prompt 1. Prompt 2 is unseen during fine-tuning and supports the cross-prompt analysis in Section~\ref{sec:crossprompt}.}
\label{tab:split}
\begin{tabular}{ccccc}
\toprule
 & \textbf{Train} & \multicolumn{2}{c}{\textbf{Test (held out)}} & \\
\cmidrule(lr){3-4}
\textbf{Score} & \textbf{P1} & \textbf{P1} & \textbf{P2} & \textbf{Test total} \\
\midrule
1.0 & 1  & 0  & 2  & 2 \\
1.5 & 3  & 0  & 2  & 2 \\
2.0 & 17 & 1  & 20 & 21 \\
2.5 & 17 & 15 & 25 & 40 \\
3.0 & 17 & 48 & 55 & 103 \\
3.5 & 17 & 23 & 45 & 68 \\
4.0 & 17 & 15 & 38 & 53 \\
4.5 & 17 & 18 & 28 & 46 \\
5.0 & 14 & 0  & 25 & 25 \\
\midrule
\textbf{Total} & \textbf{120} & \textbf{120} & \textbf{240} & \textbf{360} \\
\bottomrule
\end{tabular}
\end{table}

Two important consequences follow. Because the sparse tails are deliberately over-sampled, the training distribution is flatter than the population's. This is intentional, as a population-proportional sample would contain almost no low-scoring essays, and the model could not learn to score them. The trade-off is that the model's exposure to each score level reflects the sampling design rather than natural frequency, which is worth bearing in mind when reading the score-wise calibration in Section~\ref{sec:scorewise}. In addition, because the two prompts are cleanly separated by the split, the held-out set supports a direct test of transfer to an unseen prompt (Section~\ref{sec:crossprompt}) without any additional training. With only two prompts, however, between-prompt variance cannot be estimated and we can only say whether performance transfers to this unseen prompt.

\subsection{Model Selection}
\label{sec:model_selection}

Two instruction-tuned models were selected for fine-tuning: \texttt{google/gemma-3-27b-it} (27 billion parameters) and \texttt{meta-llama/Llama-3.3-70B-Instruct} (70 billion parameters). Instruction-tuned (``-it'' / ``Instruct'') variants were chosen over raw base models because the essay scoring task requires the model to follow a structured rubric prompt and produce formatted output (a score and optional feedback). Instruction-tuned checkpoints already possess the instruction-following behavior needed for this task structure. Fine-tuning then specializes them for rubric-aligned scoring without requiring the model to first learn how to follow instructions. Finally, both models are open-weight, as this study requires.

\subsection{Fine-Tuning with LoRA}
\label{sec:finetuning}

Fine-tuning was performed using Low-Rank Adaptation (LoRA) \citep{hu2022lora}. All base-model weights were frozen. Only the injected low-rank adapter matrices were updated. Adapters were applied to attention projections (\texttt{q\_proj}, \texttt{k\_proj}, \texttt{v\_proj}, \texttt{o\_proj}) and MLP projections (\texttt{up\_proj}, \texttt{down\_proj}, \texttt{gate\_proj}). Both models used identical LoRA and training hyperparameters to ensure a fair comparison. No quantization was applied during fine-tuning to preserve gradient precision and maintain alignment between the base weights and the learned adapters.

\begin{table}[!h]
\centering
\caption{LoRA and training hyperparameters (identical for both models).}
\label{tab:hyperparams}
\begin{tabular}{ll}
\toprule
\textbf{Parameter} & \textbf{Value} \\
\midrule
LoRA rank ($r$) & 64 \\
LoRA scaling ($\alpha$) & 128 \\
Effective scaling ($\alpha / r$) & 2 \\
LoRA dropout & 0 \\
Bias & none \\
Target modules & \texttt{q,k,v,o\_proj}; \texttt{up,down,gate\_proj} \\
\midrule
Epochs & 8 \\
Optimizer & AdamW ($\beta_1{=}0.9$, $\beta_2{=}0.999$, $\epsilon{=}10^{-8}$) \\
Learning rate & $5 \times 10^{-5}$ \\
LR scheduler & Cosine \\
Per-device batch size & 1 \\
Gradient accumulation & 8 (effective batch $\approx$ 8) \\
Precision & bf16 \\
Max sequence length & 2048 tokens \\
Seed & 42 \\
\bottomrule
\end{tabular}
\end{table}

\subsubsection{Provenance of hyperparameters and absence of test-set involvement}
\label{sec:hyperparam_provenance}

Because the validity of any held-out estimate depends on whether the held-out data influenced modeling decisions, we state the provenance of our settings explicitly. All values in Table~\ref{tab:hyperparams} were fixed before any evaluation was run. The epoch count, learning rate, scheduler, optimizer, batch configuration, and maximum sequence length were left at their default settings for LoRA supervised fine-tuning \citep{zheng2024llamafactory}. The rank and scaling factor follow the conventions established in the LoRA literature: $r = 64$ as a moderate setting and $\alpha = 2r$ \citep{hu2022lora}. No grid search, sweep, or tuning procedure was performed. The 360 held-out essays were scored once per model, after training was complete, and the resulting metrics did not feed back into any modeling decision. There is accordingly no path by which the test set could have inflated the reported estimates through hyperparameter selection.

We did not hold out a separate validation split from the 120 training essays. Training loss was logged and inspected for convergence, but this quantity is computed on the training essays themselves and is distinct from validation on held-out data. No evaluation dataset was configured during fine-tuning. Because all hyperparameters were fixed a priori (above) and the final-epoch checkpoint was used without checkpoint selection, there was no tuning, early-stopping, or model-selection decision for a validation set to inform. The training-loss curve is in any case independent of the 360 held-out test essays, so monitoring it does not affect the test-set estimates.

\subsubsection{Training infrastructure}
\label{sec:training_infra}

Fine-tuning was conducted on RunPod\footnote{\url{https://www.runpod.io}} using the LLaMA-Factory framework \citep{zheng2024llamafactory}\footnote{\url{https://github.com/hiyouga/LLaMA-Factory}} on a single NVIDIA H200 SXM GPU (141~GB VRAM). LLaMA-Factory is an open-source platform for fine-tuning large language models that unifies many of the components a researcher would otherwise have to assemble manually. It provides a single command-line and configuration-file interface to a wide range of base models (LLaMA, Gemma, Qwen, Mistral, and more than a hundred others), training methods (full fine-tuning, LoRA, QLoRA, DPO, and others), and supporting infrastructure (mixed-precision training, gradient accumulation, and dataset formatting). For this study, the framework's value lay in three properties: (1)~its YAML-based configuration files make the entire training procedure declarative and reproducible, with no custom training loop required; (2)~it handles the practical details of LoRA injection and target-module selection consistently across base models with different internal layouts, which was essential for our cross-model comparison; and (3)~it is itself open-source under the Apache 2.0 license, which preserves the end-to-end reproducibility and auditability of our pipeline. The YAML configuration files used to train both Gemma and LLaMA are released alongside the AiAWE codebase. The H200's memory capacity was sufficient to fine-tune both the 27B-parameter Gemma and the 70B-parameter LLaMA models in unquantized bf16 (16-bit) precision. Training Gemma on the $\mathcal{D}_{120}$ split completed in approximately 2 hours. LLaMA required approximately 5 hours on the same split.

\subsubsection{Input format and dual-purpose design}

Each training instance was constructed as a concatenation of a system prompt (containing the TOEFL Independent Writing rubric), a user prompt (restating the essay topic), and the essay text:
\[
x_i = \texttt{[System Prompt]} + \texttt{[User Prompt]}_i + \texttt{[Essay]}_i
\]
The system prompt embeds the full ETS scoring rubric (score levels 0--5 with descriptors). During fine-tuning, the assistant target for each instance was the bare human-assigned score (e.g., \texttt{3.5}). No JSON wrapper or \texttt{reasoning} field was provided as a target (Appendix~\ref{app:training_data}). The richer \texttt{\{score, reasoning\}} JSON output is requested only by the production prompt at inference time (Appendix~\ref{app:prompts:production}). A practical asymmetry between training and inference makes the dual-purpose behavior observable. The LoRA adapter was trained on a score-only target: the assistant output in each training instance was the bare human score, with no feedback (Appendix~\ref{app:training_data}). At inference, the production prompt requests a JSON object with both a \texttt{score} and a structured \texttt{reasoning} field across user-defined dimensions (Appendix~\ref{app:prompts:production}), and the fine-tuned model returns well-formed feedback, meaning a syntactically valid \texttt{reasoning} field, despite never having been trained on feedback examples. ``Well-formed'' here refers to structure only. The accuracy, quality, and pedagogical usefulness of that feedback are not evaluated in this study. We attribute the scoring behavior to the adapter and the feedback formatting to the preserved instruction-following of the base model. Both prompts are reproduced in Appendix~\ref{app:prompts}, and the structure of training instances is illustrated with synthetic examples in Appendix~\ref{app:training_data}.

\subsubsection{An incidental observation on LoRA rank}
\label{sec:rank_obs}

We record here an observation made during development. It is not a finding of this study, and we present it with the explicit caveat that it is unquantified. In a single exploratory run at $r = 128$, with all other settings unchanged, the LLaMA adapter stopped emitting the structured \texttt{reasoning} field required by the production prompt entirely and returned only a numeric score. Gemma, run under the same configuration, continued to produce well-formed feedback. Both models were trained at $r = 64$ for all results reported in this paper, and no scoring metrics were computed for the $r = 128$ configuration. We emphasize that this observation does not establish that LoRA hyperparameters are model-dependent in general or that LLaMA is more susceptible to capability loss than Gemma. In order to investigate this behavior in detail,  a systematic sweep with repeated runs, per-rank measurement of both scoring accuracy and feedback survival, and a quantitative criterion for what counts as capability loss would be needed (see Section \ref{sec:conclusion}) . 

\subsection{Inference Deployment}
\label{sec:inference}

At inference time, models were deployed using \texttt{llama.cpp}, an open-source C/C++ inference framework that supports quantized models via the GGUF format and enables both GPU and CPU execution. This choice was motivated by three factors: (1)~no dependency on proprietary serving infrastructure; (2)~support for tensor-level parallelism across multiple GPUs; and (3)~accessibility for researchers with consumer-grade hardware. The inference server is a local workstation equipped with an AMD Threadripper 7955WX CPU and dual NVIDIA RTX 3090 GPUs (24~GB VRAM each). To enable a fair comparison between the two models, both Gemma and LLaMA were evaluated using the same \texttt{Q4\_K\_M} (approximately 4-bit) GGUF quantization, sourced from the \texttt{bartowski} repositories on Hugging Face:

\begin{itemize}
\item \textbf{Gemma-3-27B-it} (\texttt{Q4\_K\_M}): approximately 16~GB. Fits comfortably on a single RTX 3090 alongside its LoRA adapter.
\item \textbf{LLaMA-3.3-70B-Instruct} (\texttt{Q4\_K\_M}): approximately 42.5~GB. Exceeds single-GPU capacity. Layers are distributed across both RTX 3090 GPUs using \texttt{llama.cpp}'s tensor-split functionality, utilizing the combined 48~GB of VRAM.
\end{itemize}

Briefly, \texttt{Q4\_K\_M} is a mixed-precision ``K-quant'' scheme in the GGUF format: most weights are stored at approximately 4 bits, while the token embedding matrix (which Gemma 3 also uses as its output layer) and, in about half of the transformer blocks, the attention value and feed-forward down-projections are stored at 6 bits. In the file used here, these upgrades apply in 30 of the 62 blocks, and the attention output projection is 4-bit throughout. The effective average is approximately 4.9 bits per weight (Appendix~\ref{app:quant}). It is widely used as a balance point between compression and quality \citep{kurt2026quantization}. Two properties matter for replication and are documented in Appendix~\ref{app:quant}: the per-tensor precision allocation is not fixed by the ``Q4\_K\_M'' label but depends on the quantization tool and its version, so replication should use the exact \texttt{bartowski} artifacts we specify. The choice of scheme is an engineering decision that should be evaluated against the target task and hardware rather than inherited from this study. A detailed investigation of quantization trade-offs is beyond the scope of this paper.

Using the same quantization scheme for both models means that the scoring performance differences we report in Section~\ref{sec:results} cannot be attributed to an asymmetric choice of precision. Both models operate under the same approximate bit budget per parameter. The LoRA adapters were trained on unquantized bf16 base weights but are applied at inference time on 4-bit quantized base weights. This is standard practice in open-source LLM deployment: training benefits from unquantized weights and gradients, while inference can tolerate aggressive quantization on most tasks \citep{kurt2026quantization}. The empirical results (Section~\ref{sec:results}) indicate that Q4\_K\_M is sufficient for rubric-aligned scoring in our setting.

\subsubsection{Server configuration}

Multiple \texttt{llama.cpp} server instances run concurrently on a dual-GPU setup. In production, the primary instance serves the fine-tuned Gemma model (with LoRA adapter trained on the full 480-essay dataset) for TOEFL-aligned scoring. Only Gemma is used in production, as it returned better point estimates than LLaMA on every metric in our testing and significantly lower scoring error. A separate instance runs the base Gemma model (without any LoRA adapter) as a demonstration based on a custom instructor-defined rubric for low-stakes qualitative feedback (Section~\ref{sec:awade_experimental}). Both deployed instances use the same Q4\_K\_M quantization as was used during testing. All scoring, in testing and in production (for the TOEFL rater), uses a temperature setting of 0, so that outputs are deterministic and reproducible for a given model, adapter, and prompt. Measured latency and throughput are reported in Section~\ref{sec:latency}.

\subsubsection{AWADE experimental model}
\label{sec:awade_experimental}

In addition to the fine-tuned scoring model, the AiAWE platform includes an experimental configuration that demonstrates the base Gemma model's zero-shot instruction-following capability. This mode uses the same \texttt{Q4\_K\_M} quantized Gemma model but without any LoRA adapter. Instead, a detailed custom rubric is provided entirely through the system prompt. The rubric, designed for a university-level argumentative writing course, assigns points across five categories (Introduction, First Body Paragraph, Second Body Paragraph, Conclusion, and Language \& Mechanics), and requests JSON-formatted output with both a score and paragraph-level improvement recommendations. This configuration is not evaluated against any benchmark. It is presented as a demonstration of the platform's extensibility. It shows that instructors can define their own rubrics and receive structured output without any fine-tuning. Whether the resulting scores or feedback are accurate or useful in such contexts has not been evaluated.

\subsection{Platform Implementation}
\label{sec:platform}

The scoring and fine-tuning pipeline described above constitutes the research contribution of this paper. The surrounding application, a Django web platform providing student and educator interfaces, course and user management, batch processing, custom rubric configuration, and quota management, is described in Appendix~\ref{app:platform}. It is reported there rather than here because it is engineering context for the evaluation rather than part of the evaluation itself.

\subsection{Evaluation Metrics}
\label{sec:metrics}

Model performance was evaluated on the $\mathcal{D}_{120}$ held-out set (360 essays) using the following metrics, computed against the gold-standard ETS human scores:

\begin{itemize}
\item \textbf{Root mean square error (RMSE)}: Large errors are penalized disproportionately. A single 2-point miss counts far more than four 0.5-point misses. RMSE is therefore more sensitive to occasional large deviations (outliers).

\item \textbf{Mean absolute error (MAE)}: The average of the absolute differences between prediction and truth. Every error contributes in proportion to its size, so one large miss and several small ones are treated evenhandedly.

\item \textbf{Agreement within human-rater tolerance:} The percentage of predictions falling within $\pm 0.5$ of the human score, simulating the standard inter-rater agreement threshold used in ETS's own operational scoring program.

\item \textbf{Score-wise bias:} Mean signed error per score level, revealing systematic over- or under-prediction at specific points on the rubric scale.

\item \textbf{Pearson $r$ and Spearman $\rho$:} Linear and rank-order association between model and criterion scores. Both are computed on the raw 0.5-increment values with no binning.

\item \textbf{Quadratic weighted kappa (QWK):} Chance-corrected agreement on an ordinal scale, with disagreements weighted by the square of the category distance.
\end{itemize}

\paragraph{Categorization and rounding rule.} We state the mapping explicitly, since QWK is sensitive to it. Both the ETS composite and the model predictions already lie on the 0.5-point grid and the composite is defined there. Both fine-tuned models were prompted to emit scores in 0.5 increments and did so for all 360 essays, with no off-grid values requiring correction (verified by inspection of the raw generations). The half-point grid therefore uses these nine values directly as nine equally spaced ordinal categories, with no rounding of either the predictions or the criterion. The collapsed integer grid maps both the predictions and the criterion to five categories by the same half-up rule, $c = \lfloor s + 0.5 \rfloor$ (so 3.5 becomes 4 and 4.5 becomes 5). Because the rule is applied identically to model and criterion, it changes both without advantaging either. We report QWK and exact agreement on both grids for two reasons. First, contrasting the two makes the categorization-dependence of QWK visible directly: the same predictions yield materially different coefficients under the two binnings (Section~\ref{sec:dualgrid}). Second, the integer grid yields the quantity that is arithmetically comparable to published human exact-agreement figures, which are computed on the integer scale (Section~\ref{sec:dualgrid}). We stress there that even this comparison is limited, since our criterion is an averaged composite rather than a single rater's integer score.

\paragraph{Statistical comparison of models.} Because all three systems scored the same 360 essays, the comparisons are paired: every essay yields one error from each model, and we compare those errors within essays rather than treating the three sets of scores as independent. Differences in RMSE and MAE between two models are tested with a paired-samples $t$-test on the per-essay errors (squared errors for RMSE, absolute errors for MAE). Differences in exact and within-$\pm$0.5 agreement, which are binary at the essay level, are tested with McNemar's exact test, the standard test for paired proportions, which compares only the essays the two models classify differently. QWK is a single aggregate statistic with no per-essay value and cannot be entered into a paired test. We therefore report QWK differences descriptively and do not describe them as statistically significant. A significant difference in RMSE or MAE does not establish a significant difference in QWK. We do not describe a difference as a difference on the strength of the point estimates alone.

\section{Results}
\label{sec:results}

All results reported in this section are computed on the 360-essay held-out test set ($\mathcal{D}_{120}$ split: 120 training, 360 evaluation). Both Gemma-3-27B and LLaMA-3.3-70B were fine-tuned on the same 120 essays with identical hyperparameters. The fine-tuned GPT-3.5 baseline is the model reported by \citet{wang2024effectiveness}, evaluated on the same 360 essays.

\subsection{Overall Scoring Performance}

Table~\ref{tab:overall} summarizes the global error and correlation metrics for all three models.  Paired tests for all pairwise differences are given in Table~\ref{tab:paired}.

\begin{table}[!h]
\centering
\caption{Overall scoring metrics on the 360-essay held-out test set. All values are computed by us from the per-essay predictions of all three systems. QWK is computed on the half-point grid (Section~\ref{sec:metrics}).}
\label{tab:overall}
\begin{tabular}{lcccccccc}
\toprule
\textbf{Model} & \textbf{RMSE} & \textbf{MAE} & \textbf{Pearson}~$r$ & \textbf{Spearman}~$\rho$ & \textbf{QWK} & \textbf{$\pm$0.5} & \textbf{Exact} & \textbf{Access} \\
\midrule
Gemma (LoRA)  & \textbf{0.474} & \textbf{0.335} & \textbf{0.852} & \textbf{0.845} & \textbf{0.828} & \textbf{90.56\%} & \textbf{43.33\%} & Open-weight \\
LLaMA (LoRA)  & 0.512 & 0.379 & 0.814 & 0.812 & 0.777 & 87.22\% & 37.78\% & Open-weight\\
GPT-3.5 (FT)  & 0.573 & 0.431 & 0.791 & 0.801 & 0.785 & 84.72\% & 32.78\% & Proprietary\\
\bottomrule
\end{tabular}
\end{table}

The fine-tuned Gemma model returns the most favorable point estimate on every metric. Its RMSE of 0.474 is a 17.2\% reduction relative to the GPT-3.5 baseline (0.573), and its QWK of 0.828 on the half-point grid exceeds the 0.785 we compute from the GPT-3.5 predictions. Point estimates alone cannot establish that these gaps exceed sampling error, so Table~\ref{tab:paired} reports the paired tests described in Section~\ref{sec:metrics}.

\begin{table}[!h]
\centering
\caption{Paired comparisons on the 360-essay held-out set. RMSE and MAE differences are tested with a paired-samples $t$-test on the per-essay squared and absolute errors respectively ($df = 359$). Exact and within-$\pm$0.5 agreement are compared with McNemar's exact test. QWK is a single aggregate statistic and is reported descriptively, not tested (see text). Negative RMSE/MAE differences and positive QWK and agreement differences favor the first-named model.}
\label{tab:paired}
\begin{tabular}{llrcr}
\toprule
\textbf{Comparison} & \textbf{Metric} & \textbf{Difference} & \textbf{Test} & \textbf{$p$} \\
\midrule
\multirow{5}{*}{Gemma vs.\ LLaMA}
 & RMSE            & $-0.038$   & $t = -2.39$ & $.017$ \\
 & MAE             & $-0.044$   & $t = -2.62$ & $.009$ \\
 & QWK             & $+0.051$   & descriptive & n/a \\
 & Exact           & $+5.56$ pp & McNemar     & $.072$ \\
 & Within $\pm$0.5 & $+3.33$ pp & McNemar     & $.065$ \\
\midrule
\multirow{5}{*}{Gemma vs.\ GPT-3.5}
 & RMSE            & $-0.099$    & $t = -3.85$ & $< .001$ \\
 & MAE             & $-0.096$    & $t = -4.03$ & $< .001$ \\
 & QWK             & $+0.043$    & descriptive & n/a \\
 & Exact           & $+10.56$ pp & McNemar     & $.004$ \\
 & Within $\pm$0.5 & $+5.83$ pp  & McNemar     & $.009$ \\
\midrule
\multirow{5}{*}{LLaMA vs.\ GPT-3.5}
 & RMSE            & $-0.061$   & $t = -2.39$ & $.017$ \\
 & MAE             & $-0.051$   & $t = -2.05$ & $.041$ \\
 & QWK             & $-0.008$   & descriptive & n/a \\
 & Exact           & $+5.00$ pp & McNemar     & $.200$ \\
 & Within $\pm$0.5 & $+2.50$ pp & McNemar     & $.343$ \\
\bottomrule
\end{tabular}
\end{table}

\paragraph{Gemma outperforms the GPT-3.5 baseline.} The error metrics are significant by paired $t$-test (RMSE and MAE both $p < .001$) and the agreement rates by McNemar ($p = .004$ and $.009$). The QWK difference ($+0.043$) is reported descriptively and was not tested. This comparison is the study's principal claim, and the paired tests support it.

\paragraph{Gemma outperforms LLaMA, but not on every metric.} The advantage is significant for RMSE and MAE (paired $t$-tests, $p = .017$ and $.009$). Gemma's QWK is also higher, by $0.051$, but this difference was not tested and is reported descriptively. It is not significant for exact agreement ($p = .072$) or within-$\pm$0.5 agreement ($p = .065$, McNemar), where the discordant-pair counts are modest (66 vs.\ 46, and 24 vs.\ 12, respectively). The correct summary is that Gemma achieves significantly lower scoring error than LLaMA, with a descriptively higher QWK, while the two models' agreement rates cannot be distinguished at conventional thresholds on a sample of this size.

\paragraph{LLaMA and GPT-3.5 have nearly identical QWK values.} Their QWK values differ by only $0.008$ (0.777 vs.\ 0.785). This difference was not tested, and we draw no inference from it. We note this explicitly because a gap of this size invites a substantive explanation, such as the compressed prediction range documented in Section~\ref{sec:predrange}. Although the range compression is real, it does not follow that it produced a meaningful difference in ordinal agreement. However, LLaMA does outperform GPT-3.5 on RMSE and MAE (paired $t$-tests, $p = .017$ and $.041$).

Both open-weight models therefore improve on the fine-tuned GPT-3.5 baseline in scoring error on these essays, and Gemma additionally improves on agreement. At the looser $\pm$1.0 threshold, agreement is near-ceiling for all three systems (Gemma 99.44\%, LLaMA 99.17\%, GPT-3.5 96.39\%): only 2 of Gemma's 360 predictions deviate from the composite by more than one point.

\subsection{Agreement Under Two Categorizations}
\label{sec:dualgrid}

Because QWK and exact agreement both depend on how continuous predictions are mapped to categories, Table~\ref{tab:dualgrid} reports each on the half-point grid the models predicted on and on the collapsed integer grid on which human raters operate.

\begin{table}[!h]
\centering
\caption{Agreement metrics under two categorizations of the same 360 predictions. The collapsed grid applies half-up rounding to the nearest integer to both the model score and the ETS composite.}
\label{tab:dualgrid}
\begin{tabular}{lcccc}
\toprule
 & \multicolumn{2}{c}{\textbf{QWK}} & \multicolumn{2}{c}{\textbf{Exact agreement}} \\
\cmidrule(lr){2-3}\cmidrule(lr){4-5}
\textbf{Model} & \textbf{Half-point (9 cat.)} & \textbf{Integer (5 cat.)} & \textbf{Half-point} & \textbf{Integer} \\
\midrule
Gemma   & 0.828 & 0.765 & 43.33\% & 67.22\% \\
LLaMA   & 0.777 & 0.700 & 37.78\% & 61.94\% \\
GPT-3.5 & 0.785 & 0.707 & 32.78\% & 57.50\% \\
\bottomrule
\end{tabular}
\end{table}

The two grids preserve the ordering of the systems but change every value substantially: a QWK of 0.828 and a QWK of 0.765 describe the same set of predictions. QWK figures are comparable across studies only when category count, category width, marginal distributions and rounding rule all coincide, and in the AES literature they frequently do not. Full confusion matrices for all three systems on both grids are provided in Appendix~\ref{app:confusion}.

The exact-agreement columns are relevant to one specific comparison. Published figures for human exact agreement on TOEFL Independent Writing are computed on the integer scale \citep{wang2014monitoring}. On that scale Gemma reproduces the rounded ETS composite in 67.22\% of cases. We report this because it is the arithmetically comparable quantity, but we do not offer it as evidence of human parity. As set out in Section~\ref{sec:dataset}, the ETS composite is a smoothed average of two or more ratings and carries lower error variance than any single rating, so agreement with it is not the same quantity as agreement between two independent raters. A properly grounded human--model comparison would require the individual rater scores, which are not distributed with this dataset (Section~\ref{sec:limitations}).

\subsection{Score-Wise Error Analysis}
\label{sec:scorewise}

To understand model behavior across the rubric spectrum, we computed RMSE and mean signed bias at each ETS score level. The results are shown in Table~\ref{tab:scorewise}, and several patterns emerge from the score-wise analysis.

\begin{table}[!h]
\centering
\caption{Score-wise error analysis. Bias is defined as $\hat{y} - y$ (positive = overprediction).}
\label{tab:scorewise}
\begin{tabular}{c|ccc|ccc}
\toprule
 & \multicolumn{3}{c|}{\textbf{Gemma}} & \multicolumn{3}{c}{\textbf{LLaMA}} \\
\textbf{ETS Score} & \textbf{$n$} & \textbf{RMSE} & \textbf{Bias} & \textbf{$n$} & \textbf{RMSE} & \textbf{Bias} \\
\midrule
1.0 & 2  & 0.500 & +0.50 & 2  & 1.000 & +1.00 \\
1.5 & 2  & 0.354 & +0.25 & 2  & 0.500 & +0.50 \\
2.0 & 21 & 0.362 & +0.17 & 21 & 0.423 & +0.31 \\
2.5 & 40 & 0.296 & +0.08 & 40 & 0.411 & +0.09 \\
3.0 & 103 & 0.388 & $-$0.13 & 103 & 0.415 & $-$0.04 \\
3.5 & 68 & 0.454 & $-$0.19 & 68 & 0.450 & $-$0.11 \\
4.0 & 53 & 0.571 & $-$0.29 & 53 & 0.536 & $-$0.35 \\
4.5 & 46 & 0.585 & $-$0.42 & 46 & 0.647 & $-$0.58 \\
5.0 & 25 & 0.671 & $-$0.46 & 25 & 0.781 & $-$0.70 \\
\bottomrule
\end{tabular}
\end{table}

\paragraph{Regression toward the mean.} Both models show the pattern expected of an imperfect predictor of a bounded outcome: they overpredict low scores (positive bias at 1.0--2.5) and underpredict high scores (negative bias at 3.5--5.0). This ``regression to the mean'' is common in AES systems and is exacerbated by class imbalance: scores of 1.0 and 1.5 together account for only 4 of the 360 test essays, while the 3.0 bin alone contains 103.

\paragraph{Gemma shows better calibration.} Gemma's bias progression is smoother and more gradual across the score range (from +0.50 at score 1.0 to $-$0.46 at score 5.0). LLaMA shows sharper swings, particularly at the extremes: it overpredicts score-1.0 essays by a full point (+1.00 bias) and underpredicts score-5.0 essays by 0.70 points. This suggests that Gemma has better internalized the full rubric scale.

\paragraph{Mid-range accuracy is strong.} Both models perform best in the 2.0--3.5 range, which is also the most densely populated region of the dataset. Gemma achieves its lowest RMSE (0.296) at score 2.5, while both models remain below 0.46 RMSE from 2.0 to 3.5.

\paragraph{High scores are the hardest to predict.} Error increases monotonically above score 3.0 for both models. At score 5.0, Gemma's RMSE is 0.671 and LLaMA's is 0.781, with negative bias throughout the upper range ($-0.46$ and $-0.70$ respectively at score 5.0). Both the direction and the location of this effect are consistent with the wider literature: in a scoping review of 26 studies of GPT-based ratings of EFL writing, \citet{chen2026evaluating} reports that LLM-based raters tend to score more severely than human raters. In the present data, this severity is concentrated at the top of the scale. Two mechanisms are plausible and are not mutually exclusive: the regression-toward-the-mean effect described above, and a genuine reluctance of instruction-tuned models to award the maximum score, which would be a property of the base models' alignment rather than of our adaptation. The present design cannot separate the two, since doing so would require either a balanced training distribution at the tails or a comparison against a model with the same score prior but different alignment. We therefore treat this as an open question.

\subsection{Prediction Range and Score Compression}
\label{sec:predrange}

An important behavioral difference between the models is the range of their predictions. The ETS ground truth spans 1.0 to 5.0. Gemma's predictions range from 1.5 to 5.0, so it never assigns a score of 1.0. LLaMA's range is 2.0 to 5.0, missing the lowest two score levels entirely. Table~\ref{tab:pred_dist} shows the prediction frequency distributions.

\begin{table}[!h]
\centering
\caption{Prediction frequency distribution by model, against the test-set ground-truth distribution ($N = 360$).}
\label{tab:pred_dist}
\begin{tabular}{cccc}
\toprule
\textbf{Score} & \textbf{ETS} & \textbf{Gemma} & \textbf{LLaMA} \\
\midrule
1.0 & 2   & 0  & 0  \\
1.5 & 2   & 4  & 0  \\
2.0 & 21  & 21 & 25 \\
2.5 & 40  & 72 & 63 \\
3.0 & 103 & 98 & 89 \\
3.5 & 68  & 72 & 80 \\
4.0 & 53  & 50 & 85 \\
4.5 & 46  & 31 & 16 \\
5.0 & 25  & 12 & 2  \\
\midrule
\textbf{Total} & \textbf{360} & \textbf{360} & \textbf{360} \\
\bottomrule
\end{tabular}
\end{table}

LLaMA's prediction distribution reveals a notable compression of the upper score range: it assigns 85 essays a score of 4.0 but only 16 at 4.5 and just 2 at 5.0, despite the ground truth containing 46 and 25 essays at those levels respectively. This compression, combined with LLaMA's inability to predict scores below 2.0, suggests that LLaMA's LoRA adaptation has effectively narrowed its scoring range to approximately 2.0--4.0, with predictions at 4.5 and 5.0 being rare. Gemma maintains a broader and more balanced prediction distribution that more closely mirrors the ground truth.

\subsection{Cross-Prompt Generalization}
\label{sec:crossprompt}

Because all 120 training essays were drawn from prompt 1 (Section~\ref{sec:split}), the held-out set separates cleanly into 120 essays on the training topic and 240 essays on a topic neither model observed during fine-tuning. This supports a direct test of whether scoring accuracy transfers to an unseen prompt. Table~\ref{tab:crossprompt} reports performance on each partition.

\begin{table}[!h]
\centering
\caption{Scoring performance by prompt. P1 is the topic from which the training essays were drawn. P2 was not seen during fine-tuning. The final column reports an independent-samples (Welch) $t$-test comparing the per-essay absolute errors on the unseen prompt against the seen prompt. A positive $t$ indicates worse performance on the unseen prompt. RMSE, MAE, QWK and $\pm$0.5 are descriptive per-prompt values.}
\label{tab:crossprompt}
\begin{tabular}{llcccccl}
\toprule
\textbf{Model} & \textbf{Prompt} & \textbf{$n$} & \textbf{RMSE} & \textbf{MAE} & \textbf{QWK} & \textbf{$\pm$0.5} & \textbf{Welch $t$ ($p$)} \\
\midrule
\multirow{2}{*}{Gemma}
 & P1 (seen)   & 120 & 0.438 & 0.317 & 0.772 & 93.33\% & \multirow{2}{*}{$+0.76$ ($.45$)} \\
 & P2 (unseen) & 240 & 0.491 & 0.344 & 0.843 & 89.17\% & \\
\midrule
\multirow{2}{*}{LLaMA}
 & P1 (seen)   & 120 & 0.477 & 0.363 & 0.711 & 90.83\% & \multirow{2}{*}{$+0.68$ ($.50$)} \\
 & P2 (unseen) & 240 & 0.528 & 0.388 & 0.795 & 85.42\% & \\
\midrule
\multirow{2}{*}{GPT-3.5}
 & P1 (seen)   & 120 & 0.629 & 0.492 & 0.661 & 80.00\% & \multirow{2}{*}{$-2.13$ ($.034$)} \\
 & P2 (unseen) & 240 & 0.542 & 0.400 & 0.827 & 87.08\% & \\
\bottomrule
\end{tabular}
\end{table}

Neither open-weight model degrades significantly on the unseen prompt: an independent-samples $t$-test on the per-essay errors finds no significant difference between the seen and unseen prompts for Gemma ($t = 0.76$, $p = .45$) or LLaMA ($t = 0.68$, $p = .50$), even though the point-estimate RMSE rises slightly (0.438 to 0.491, and 0.477 to 0.528, respectively). The fine-tuned GPT-3.5 baseline shows the opposite pattern, scoring somewhat more accurately on the unseen prompt ($t = -2.13$, $p = .034$). This concerns the baseline rather than our open-weight system and does not affect the claim. A non-significant difference does not by itself show that accuracy was unchanged. We therefore conclude only that neither model showed a detectable loss of accuracy on this unseen prompt, not that its accuracy was equivalent.

This result is consistent with the prompt-robustness finding that \citet{wang2024effectiveness} reported for fine-tuned GPT-3.5, which that study tested more thoroughly than we can here, having trained separate models on prompt 1, prompt 2, and both. As separate, corroborating evidence, a companion evaluation \citep{gayed2026l1bias} reports stable accuracy for the production model across eight TOEFL11 prompts that are disjoint from the two used here. That evaluation concerns a different model from the one tested here (the production adapter, trained on all 480 essays; Section~\ref{sec:inference}). This conclusion has three limits. First, with only two prompts in this dataset we can establish that performance transferred to this unseen prompt, but we cannot estimate between-prompt variance, which would require several prompts. Second, we did not train a mirrored model on prompt 2 and evaluate it on prompt 1, which would test the symmetry of the effect. Third, both prompts are TOEFL Independent Writing tasks, so the test concerns transfer within a single task family.

Table~\ref{tab:crossprompt} also illustrates the categorization dependence discussed in Section~\ref{sec:dualgrid}. Gemma's QWK is higher on the unseen prompt (0.843) than on the seen one (0.772), while its RMSE is simultaneously worse. The two results are compatible because the two partitions have different score marginals. The unseen partition spans the full 1.0–5.0 range while the seen partition is compressed to 2.0–4.5, and QWK rewards correct ordering across a wider range.

\subsection{Inference Latency and Throughput}
\label{sec:latency}

We report timings from a dedicated measurement run over the same 360 held-out essays used for the accuracy evaluation, on the same \texttt{bartowski} Gemma-3-27B-it \texttt{Q4\_K\_M} artifact, so that the measured workload and the evaluated workload coincide (Table~\ref{tab:latency}, protocol in Appendix~\ref{app:bench}). The system emits its first token 1.28 s after a request (p95 1.63 s) and completes a full score with rubric-referenced feedback in a median of 14.23 s (p95 16.61 s). The difference between the two reflects output length rather than system speed: the feedback runs to a median of 419 tokens, generated at about 32 tokens per second, while prefill of the roughly 1{,}100-token prompt adds about 1.3 s. Generation throughput is stable across the test set (5th and 95th percentiles of the decode rate 32.2 and 32.5 tokens per second), so end-to-end latency is essentially a function of feedback length and can be estimated from it for capacity planning.

\begin{table}[!h]
\centering
\caption{Measured inference timings over the 360 held-out essays. Gemma-3-27B-it, \texttt{bartowski} \texttt{Q4\_K\_M} GGUF, single RTX 3090, \texttt{llama.cpp} with a 4096-token context, requests issued sequentially. $n = 360$.}
\label{tab:latency}
\begin{tabular}{lr}
\toprule
\textbf{Measure} & \textbf{Value} \\
\midrule
Prompt tokens (median [min, max]) & 1120 [530, 2185] \\
Generated tokens (median [min, max]) & 419 [288, 611] \\
Prefill throughput (median) & 898 tokens/s \\
Generation throughput (median) & 32.4 tokens/s \\
\midrule
Time to first token, p50 & 1.28 s \\
Time to first token, p95 & 1.63 s \\
End-to-end latency, p50 & 14.23 s \\
End-to-end latency, p95 & 16.61 s \\
End-to-end latency, mean (max) & 14.26 s (20.34 s) \\
\midrule
Sequential throughput & 4.0 essays/min \\
\bottomrule
\end{tabular}
\end{table}

These figures are relevant to two intended modes of use. For an individual student, a first token after about a second permits a streamed, responsive interface. For batch scoring, the sequential rate of 4.0 essays per minute clears a thirty-essay class set in under eight minutes, and the platform's batch interface (Appendix~\ref{app:platform}) processes larger submissions offline. The measurements therefore establish technical feasibility for both modes on a single consumer GPU under the tested conditions, with one request at a time and no concurrent users. They do not establish performance under concurrent load, reliability in sustained operational use, or suitability for classroom use more broadly (Section~\ref{sec:rq4}).

\section{Discussion}
\label{sec:discussion}

This study set out to answer four research questions about the development of an open-source automated writing evaluation system. We address each in turn, drawing on the results in Section~\ref{sec:results}.

\subsection{RQ1: Can a Locally Deployable Open-Weight Model Match the Fine-Tuned GPT-3.5 Baseline?}

The results show that, in this setting, a locally deployable open-weight model adapted with LoRA on modest training data can achieve scoring performance competitive with, and on these data significantly better than, the fine-tuned GPT-3.5 baseline. On the same 360-essay held-out test set, the fine-tuned Gemma-3-27B model achieves an RMSE of 0.474, a QWK of 0.828, and an agreement rate of 90.56\% within $\pm$0.5 of the human score. These numbers surpass the corresponding values for the fine-tuned GPT-3.5 baseline of \citet{wang2024effectiveness}, recomputed from its per-essay predictions on the same essays (RMSE 0.573, QWK 0.785, agreement 84.72\%). The fine-tuned LLaMA-3.3-70B also outperforms the GPT-3.5 baseline on RMSE and MAE, while their QWK values are nearly identical (0.777 vs.\ 0.785).

The practical significance of this result lies in the concerns set out in Section~\ref{sec:Intro}. Competitive LLM-based AES results have come predominantly from proprietary APIs, which raises data-sovereignty, reproducibility, and cost concerns. Our results show that, for this task and dataset, a self-hosted open-weight model can reach scoring performance competitive with a fine-tuned proprietary baseline while keeping student data local and the pipeline auditable. This is a demonstration of feasibility, not a general comparison of open-weight and proprietary models as classes. We compared two open-weight models with one fine-tuned GPT-3.5 model on one dataset under one training regime, and more recent proprietary models were not evaluated.

A comparison is also possible with the systematic review by \citet{emirtekin2025large}, in which the QWK values reported across the reviewed studies span roughly 0.47 to 0.99, with a median near 0.79. Our Gemma QWK of 0.828 falls above that median, in the upper half of the distribution but well short of its ceiling. This placement should not be read too strongly in either direction: the reviewed studies differ in task, rubric, category count and rounding rule, and for the reasons set out in Section~\ref{sec:dualgrid} QWK values are not straightforwardly comparable across such differences. Nevertheless, the comparison indicates that our system is not an outlier in the literature. 

\subsection{RQ2: How Do the Two LoRA-Adapted Configurations Compare?}

Under identical training data (120 essays) and identical LoRA hyperparameters, the 27B-parameter Gemma configuration returns better point estimates than the 70B-parameter LLaMA configuration on every metric we computed, and the advantage is statistically supported on the error metrics: RMSE (0.474 vs.\ 0.512, $p = .017$) and MAE (0.335 vs.\ 0.379, $p = .009$, paired $t$-tests). Its QWK is also higher (0.828 vs.\ 0.777), a difference we report descriptively because it was not tested. The differences in exact agreement (43.33\% vs.\ 37.78\%, $p = .072$) and within-$\pm$0.5 agreement (90.56\% vs.\ 87.22\%, $p = .065$, McNemar) do not reach significance on this sample, and we do not treat them as evidence. Gemma is also better calibrated across the score range, showing a smoother bias progression, while LLaMA exhibits a compressed prediction distribution: it never predicts below 2.0 and produces only 2 predictions at score 5.0. The claim we can defend is therefore that the Gemma configuration achieved significantly lower scoring error, with a descriptively higher QWK, not that it was superior on every measure. We note also that this comparison rests on a single training run per model at a single seed. A difference of this magnitude, significant though it is under a paired test on these 360 essays, could plausibly be attenuated or reversed by seed variation, and our design cannot rule that out (Section~\ref{sec:limitations}).

This comparison does not isolate model scale. Gemma-3-27B and LLaMA-3.3-70B differ in parameter count but also in model family, pre-training corpora, instruction-tuning recipes, and internal representations, and any of these differences may account for the result. For example, they may make Gemma more receptive to rubric-based supervision in the low-data regime we operate in. A second possibility draws on \citet{biderman2024lora}: LoRA's effectiveness depends on the intrinsic rank of the task relative to the capacity of the adapter, so a single shared configuration need not be equally suitable for both models. Separating scale from family would require models of different sizes within the same family, ideally replicated across several families, which this study does not include. We can rule out one alternative explanation: both models were evaluated with the same Q4\_K\_M quantization, so the performance gap is not an artifact of a difference in inference precision between them.

The practical implication is therefore limited but useful. A larger parameter count should not by itself be assumed to guarantee better AWE scoring performance, as the two configurations examined here illustrate: under identical training and inference conditions, the 27B configuration outperformed the 70B one. For practitioners choosing among open-weight models, candidates are worth comparing empirically on the target task rather than selected by size alone. The result does not show that model scale is, in general, an unreliable predictor of downstream performance.

\subsection{RQ3: Does Scoring Transfer to an Unseen Prompt?}

Neither open-weight model showed evidence of degradation on the unseen prompt, although the evidence is bounded. Fine-tuning drew exclusively on essays written to prompt 1, yet neither model's error increased significantly on the 240 essays written to prompt 2 (Gemma $t = 0.76$, $p = .45$; LLaMA $t = 0.68$, $p = .50$; independent-samples $t$-tests on the per-essay errors). This indicates the absence of a detectable loss rather than evidence of equivalent accuracy.

This is relevant to the practical value of such systems. A scorer that required rubric-graded training data for every new prompt would be of limited use to an instructor, since the labeling cost would recur with each assignment. The present result is consistent with, but does not establish, the absence of that requirement for this task, since the adapter's error did not detectably increase on the second prompt. It is consistent with the prompt-robustness result of \citet{wang2024effectiveness} for fine-tuned GPT-3.5, and with the multi-prompt evaluation of the production model in \citet{gayed2026l1bias}, although the latter concerns a different model. The present data alone do not establish what the adapter has learned. Whether it is closer to a mapping from rubric criteria to scores than to prompt-specific content remains a hypothesis that more prompts would be needed to test.

We are nonetheless cautious about how far to generalize. Both prompts here are TOEFL Independent Writing tasks: argumentative essays, written under timed test conditions, by adult L2 writers, on everyday topics. Transfer between two prompts drawn from the same task family is a considerably weaker test than transfer across genres or writing conditions, and the present data say nothing about the latter. With only two prompts we also cannot estimate between-prompt variance, only detect whether this particular transfer degraded performance.

\subsection{LoRA Rank and Capability Retention}
\label{sec:forgetting_discussion}

The rank observation in Section~\ref{sec:rank_obs} points to a deployment issue that holds independently of that single run. The dual-purpose behavior on which this system depends is not guaranteed by the training objective: the adapter is trained on a score-only target, and the model's ability to produce rubric-referenced feedback is inherited from the base model's instruction-following, which LoRA preserves only to a degree \citep{hu2022lora, biderman2024lora, rathore2025rank}. The consequence for AWE is that a configuration change which improves scoring accuracy can silently remove the feedback capability, because the training loss never measures feedback quality and nothing in a training curve or held-out RMSE would reveal it. Capability retention should therefore be treated as an explicit acceptance criterion, verified separately from the scoring objective at every configuration change (for example, checking that the response still parses as JSON with a non-empty \texttt{reasoning} field). Whether architectures differ systematically in their susceptibility remains open. The systematic rank sweep that would settle it is described in Section~\ref{sec:conclusion}.

\subsection{RQ4: Is Local Deployment Technically Feasible on Consumer-Grade Hardware?}
\label{sec:rq4}

The fourth question concerns technical feasibility: whether the fine-tuned scoring model can run on consumer-grade hardware, using open-source infrastructure, at practically usable latency and throughput. The measured answer is yes, under the tested conditions and with a qualification about what ``fast enough'' means. The deployed Gemma model runs on a single NVIDIA RTX 3090 (24 GB VRAM) using \texttt{llama.cpp} and Q4\_K\_M quantization (approximately 16 GB). During accuracy testing, the LLaMA model was evaluated at the same Q4\_K\_M quantization, distributed across two RTX 3090s via tensor-split parallelism. Measured timings over the 360 held-out essays are reported in Section~\ref{sec:latency}, with the protocol in Appendix~\ref{app:bench}: a complete score with rubric-referenced feedback takes 14.23~s at the median, the first token is emitted after 1.28~s, and sequential throughput is 4.0 essays per minute. The qualification is that these two latency figures answer different questions. Against the first, the system is interactive. Against the second, it is not instantaneous, and an educator scoring a class set of thirty essays should budget around seven and a half minutes. These measurements establish technical feasibility on the tested hardware. They do not establish classroom suitability in a broader sense, which would also depend on performance under concurrent use, usability, workflow integration, reliability under authentic conditions, and the pedagogical usefulness of the output (Section~\ref{sec:limitations}).

The scoring results in Section~\ref{sec:results} were obtained with the adapter applied to 4-bit quantized base weights, despite the train-inference precision mismatch described in Section~\ref{sec:inference}. Q4\_K\_M was therefore sufficient to reach these results. Because we did not score the same essays at higher precision, however, we cannot say how much accuracy, if any, the quantization cost (Section~\ref{sec:limitations}). This outcome is broadly consistent with the quantization evaluations reported by \citet{huang2024empirical} and \citet{kurt2026quantization}, which find that 4-bit schemes introduce measurable but often acceptable degradation on tasks that do not require fine-grained numerical reasoning. 

More broadly, the choice of \texttt{llama.cpp} as the inference runtime was deliberate. Unlike vendor-specific serving frameworks, \texttt{llama.cpp} supports both GPU and CPU execution and imposes no proprietary dependencies. This means that the released adapters can be served by researchers and educators with access to a single consumer GPU, or a CPU-only environment for smaller models. Repeating the fine-tuning requires more GPU memory than consumer hardware provides. We used a single rented H200 GPU for approximately 2 hours (Gemma) and 5 hours (LLaMA) (Section~\ref{sec:training_infra}). This lowers the barrier to entry for LLM-based AES research.

\subsection{Limitations}
\label{sec:limitations}

\paragraph{Single dataset, single genre.} All evaluation is conducted on 480 TOEFL Independent Writing essays responding to two prompts. While the dataset is professionally scored and widely used, it represents a narrow slice of real-world writing: argumentative essays from adult L2 English learners under test conditions. Performance on other genres (narrative, expository, research writing), other populations (K--12, L1 English writers, learners of other languages), or longer-form compositions (term papers, theses) is not established.

\paragraph{Class imbalance at the extremes.} The dataset contains very few essays at the lowest (1.0, 1.5) and highest (5.0) score levels. Scores below 2.0 account for only 8 essays and scores of 5.0 account for 39. The score-wise error analysis shows that both models struggle at these extremes, and the small sample sizes in each bin make the reported bin-level RMSE and bias estimates noisy. More data at the tails would both improve model performance and yield tighter estimates of extreme-score behavior.

\paragraph{Training split reporting.} The quantitative results reported in Section~\ref{sec:results} come from the 120-train/360-test split, which provides an honest generalization estimate. The production model deployed on the AiAWE platform is trained on all 480 essays for maximum use of available supervision, but this production model has not been evaluated on a separate held-out set because none exists within this dataset. We therefore cannot report point estimates of scoring quality for the production model. We can only report them for the 120-split variant.

\paragraph{Train-inference quantization mismatch.} LoRA adapters were trained with unquantized bf16 base weights but applied at inference time on Q4\_K\_M (4-bit) quantized base weights. The empirical results suggest this mismatch is tolerable for rubric-aligned scoring at the bit-widths we used, but we did not compare scoring quality across different quantization schemes. It is possible that a higher-bit quantization (e.g., Q6\_K or Q8\_0) or an unquantized bf16 Gemma inference run would achieve slightly better numbers than the ones we report, and that a more aggressive quantization (e.g., Q3 or lower) would begin to degrade performance. We discuss other quantization options and the factors that practitioners should consider when choosing a scheme in Appendix~\ref{app:quant}.

\paragraph{The rank observation is anecdotal and is not a finding.} The observation that LLaMA ceased producing feedback at $r = 128$ derives from a single exploratory run, with no repetition, no intermediate rank settings, no quantitative criterion for capability loss, and no scoring metrics computed for that configuration. It cannot support a claim about architectural differences in rank tolerance, and we advance none. It is reported in Section~\ref{sec:rank_obs} only because other researchers using LoRA adapters for AWE should be aware of it.

\paragraph{No fairness analysis.} This study does not itself analyze scoring behavior across L1 groups or other learner characteristics. Prior work \cite{liu2025enhancing} has shown that fine-tuned models can exhibit differential performance across L1 backgrounds, and deployment in high-stakes settings would require a fairness audit that this study does not provide. A companion evaluation of the production model \citep{gayed2026l1bias} finds a first-language-linked scoring offset that aggregate agreement statistics of the kind reported here cannot detect. A full fairness analysis is therefore an essential direction for future work.

\paragraph{Intended use is formative and low-stakes.} We position AiAWE for formative, low-stakes use, and we do not claim it is suitable for placement, admissions, certification, or any decision with material consequences for a student. Following \citet{williamson2012framework} and \citet{ferrara2022validity}, the evidence assembled here supports a validity argument for a narrow interpretation and use of the scores, not a general warrant for them, and it says nothing directly about the generated feedback. A high-stakes deployment would require, at minimum, a completed subgroup fairness audit, evidence of construct relevance beyond agreement statistics, and a monitoring plan for score drift over time. None of those are provided by this study.

\paragraph{No rater-level data.} The dataset as released to approved researchers contains only the final ETS composite score. Individual rater scores are not distributed, so we cannot report model agreement against each rater separately, cannot compute human--human agreement on these 360 essays, and cannot decompose model error into components attributable to rater disagreement versus model error. This constrains our comparisons to a smoothed multi-rater composite, a point we highlight wherever agreement figures are reported.

\paragraph{No evaluation of feedback or classroom use.} The AiAWE platform has been deployed and used by students and instructors, but this study evaluates only its scoring component, against the ETS criterion. We did not evaluate the accuracy, quality, or usefulness of the generated qualitative feedback, and we do not report classroom outcomes, formative-feedback efficacy, or pedagogical impact. Nothing reported here should be read as validating the feedback. Because the adapter was trained on a score-only target (Section~\ref{sec:finetuning}), the feedback is produced by the base model's instruction-following rather than learned from rubric-referenced examples, which is a further reason not to extend the scoring evidence to it.

\section{Conclusion}
\label{sec:conclusion}

This study developed and evaluated AiAWE, an open-source automated writing evaluation system that uses a LoRA-adapted Gemma-3-27B model to score argumentative essays on the TOEFL Independent Writing rubric. Fine-tuning was performed on 120 human-scored essays from the ETS dataset. Evaluation was performed on the remaining 360. The fine-tuned Gemma model achieved an RMSE of 0.474, a QWK of 0.828, and an agreement rate of 90.56\% within $\pm$0.5 of the human score, with significantly lower scoring error than both the fine-tuned LLaMA-3.3-70B configuration and the fine-tuned GPT-3.5 baseline reported by \citet{wang2024effectiveness} on the same dataset.

Four findings emerge from this work, each bounded by the conditions examined. First, a locally deployable open-weight model adapted with LoRA achieved scoring performance competitive with, and on these data better than, the fine-tuned GPT-3.5 baseline: Gemma improved significantly on it on every metric tested under paired comparison. This demonstrates that competitive scoring is feasible in this setting without sending student writing to a commercial API. It is not a general comparison of open-weight and proprietary models. Second, in a comparison of two specific configurations, the 27B Gemma achieved significantly lower scoring error than the 70B LLaMA, with a descriptively higher QWK, under identical training conditions, though their agreement rates could not be distinguished on this sample. Because size is confounded with model family, this does not isolate the effect of scale, but it does illustrate that a larger parameter count should not by itself be assumed to guarantee better AWE scoring performance. Third, when the models adapted on essays from one prompt were applied to one unseen prompt from the same task family, neither showed a statistically detectable increase in error, although this does not establish that accuracy was unchanged. Fourth, the LoRA-adapted 27B model ran on a single consumer GPU using open-source inference tools, at a 4-bit quantization level that achieved the scoring results reported here, with a median of 14.2~s per complete response. This establishes technical feasibility under the tested conditions, not classroom suitability more broadly.

The AiAWE platform (\url{https://app.awade.gec.waseda.ac.jp}) operationalizes these findings as a publicly accessible web application, with a student-facing interface that returns scores and generated feedback, and an educator-facing interface for class management, batch scoring, and custom rubric configuration. We present it as a working demonstration that the pipeline is deployable on accessible hardware. The evidence reported here concerns its scoring component. The generated feedback has not been evaluated, and the platform is offered for formative, low-stakes use on that basis.

Several directions would extend this work.  A companion evaluation reported in \citet{gayed2026l1bias} identifies a first-language-linked scoring offset but does not establish its mechanism, and the questions that follow are consequential for deployment. Chief among them is whether the offset originates in the base model's pre-training, in the composition of the fine-tuning data, or in the rubric's interaction with particular rhetorical conventions. Evaluating the system on additional genres and datasets, including narrative writing, expository prose, research abstracts, K--12 essays and non-English writing, would establish how far the LoRA adaptation approach generalizes beyond argumentative TOEFL essays. The cross-prompt result reported in Section~\ref{sec:crossprompt} shows transfer between two prompts within a single task family, which is a considerably weaker claim than transfer across genres.

A quantitative characterization of LoRA rank sensitivity across model families would provide practical guidance for the dual-purpose fine-tuning pattern used here. This would involve systematic rank sweeps ($r \in \{8, 16, 32, 64, 128, 256\}$) on multiple base models, measuring both scoring accuracy and the survival of qualitative feedback generation at each setting. Such an experiment would convert the qualitative observation reported in this study into replicable design guidance for AES practitioners. Distillation to smaller models is another natural extension. The Gemma-3-27B model, while deployable on a single 3090, may still be heavy for real-time classroom use at scale. Distilling its LoRA-adapted scoring and feedback behaviors into a smaller (e.g., 7B or 4B) student model could extend deployment to lower-cost GPUs or even CPU-only environments, making the system accessible to more institutions.

A direct evaluation of the generated feedback is a prerequisite for any claim about it: its accuracy with respect to the rubric, its agreement with expert feedback on the same essays, and its usefulness to learners. Finally, a classroom study is needed to measure the pedagogical impact of the generated feedback on revision quality, student engagement, and writing development over time. This would move the evaluation beyond scoring accuracy and into the educational outcomes that ultimately justify the deployment of AWE systems. Such a study would also provide the fairness audit that a full-scale deployment requires, examining whether model scores and feedback vary systematically across student subgroups. The open-source architecture of AiAWE is well suited to supporting this kind of research, because researchers can inspect, modify, and extend every component of the pipeline.

\bibliographystyle{apalike}
\bibliography{references_final}

@article{atkinson2025hybrid,
  title={An {LLM}-based hybrid approach for enhanced automated essay scoring},
  author={Atkinson, John and Palma, Diego},
  journal={Scientific Reports},
  volume={15},
  number={1},
  pages={Article 14551},
  year={2025},
  doi={10.1038/s41598-025-87862-3}
}

@article{biderman2024lora,
  title={{LoRA} learns less and forgets less},
  author={Biderman, Dan and Portes, Jacob and Gonzalez Ortiz, Jose Javier and Paul, Mansheej and Greengard, Philip and Jennings, Connor and King, Daniel and Havens, Sam and Chiley, Vitaliy and Frankle, Jonathan and Blakeney, Cody and Cunningham, John P.},
  journal={Transactions on Machine Learning Research},
  year={2024},
  url={https://openreview.net/forum?id=aloEru2qCG}
}

@article{blanchard2013toefl11,
  title={{TOEFL11}: A corpus of non-native {English}},
  author={Blanchard, Daniel and Tetreault, Joel and Higgins, Derrick and Cahill, Aoife and Chodorow, Martin},
  journal={ETS Research Report Series},
  volume={2013},
  number={2},
  year={2013},
  doi={10.1002/j.2333-8504.2013.tb02331.x}
}

@article{chen2026evaluating,
  title={Evaluating {GPT} ratings of {EFL} writing: A scoping review},
  author={Chen, Yi},
  journal={Assessing Writing},
  volume={68},
  pages={Article 101044},
  year={2026},
  doi={10.1016/j.asw.2026.101044}
}

@inproceedings{dettmers2024qlora,
  title={{QLoRA}: Efficient finetuning of quantized {LLMs}},
  author={Dettmers, Tim and Pagnoni, Artidoro and Holtzman, Ari and Zettlemoyer, Luke},
  booktitle={Advances in Neural Information Processing Systems 36},
  pages={10088--10115},
  year={2023},
  publisher={Neural Information Processing Systems Foundation},
  doi={10.52202/075280-0441}
}

@inproceedings{diez2024web,
  title={A web application for a cost-effective fine-tuning of open-source {LLMs} in education},
  author={Diez-Rozas, Victor and Estevez-Ayres, Iria and Alario-Hoyos, Carlos and Callejo, Patricia and Delgado Kloos, Carlos},
  booktitle={Artificial Intelligence in Education: Posters and late breaking results, workshops and tutorials, industry and innovation tracks, practitioners, doctoral consortium and blue sky},
  pages={267--274},
  year={2024},
  publisher={Springer},
  doi={10.1007/978-3-031-64312-5_32}
}

@article{emirtekin2025large,
  title={Large language model-powered automated assessment: A systematic review},
  author={Emirtekin, Emrah},
  journal={Applied Sciences},
  volume={15},
  number={10},
  pages={Article 5683},
  year={2025},
  doi={10.3390/app15105683}
}

@article{ferrara2022validity,
  title={Validity arguments for {AI}-based automated scores: Essay scoring as an illustration},
  author={Ferrara, Steve and Qunbar, Saed},
  journal={Journal of Educational Measurement},
  volume={59},
  number={3},
  pages={288--313},
  year={2022},
  doi={10.1111/jedm.12333}
}

@article{gao2025enhancing,
  title={Enhancing privacy-preserving deployable large language models for perioperative complication detection: A targeted strategy with {LoRA} fine-tuning},
  author={Gao, Shaowei and Zhao, Xu and Chen, Lihui and Yu, Junrong and Tian, Shuning and Zhou, Huaqiang and Chen, Jingru and Long, Sizhe and He, Qiulan and Feng, Xia},
  journal={npj Digital Medicine},
  volume={8},
  number={1},
  pages={Article 773},
  year={2025},
  doi={10.1038/s41746-025-02139-3}
}

@misc{jgayed_gemmalora120,
  author={Gayed, John Maurice},
  title={{gemmalora120}: {LoRA} adapter for {Gemma}-3-27{B}-it (120-sample split) [Computer software]},
  howpublished={Hugging Face. \url{https://huggingface.co/jgayed/gemmalora120}},
  year={2025}
}

@misc{jgayed_gemmalorafull,
  author={Gayed, John Maurice},
  title={{gemmalorafull}: {LoRA} adapter for {Gemma}-3-27{B}-it (full 480-sample set) [Computer software]},
  howpublished={Hugging Face. \url{https://huggingface.co/jgayed/gemmalorafull}},
  year={2025}
}

@misc{gayed2026l1bias,
  title={Investigating first-language bias in {LLM}-based automated essay scoring: A cross-prompt evaluation of an open-weight {AI}-model on {TOEFL} essays [Preprint]},
  author={Gayed, John Maurice},
  howpublished={arXiv},
  year={2026},
  doi={10.48550/arXiv.2607.14605}
}

@inproceedings{hu2022lora,
  title={{LoRA}: Low-rank adaptation of large language models},
  author={Hu, Edward J. and Shen, Yelong and Wallis, Phillip and Allen-Zhu, Zeyuan and Li, Yuanzhi and Wang, Shean and Wang, Lu and Chen, Weizhu},
  booktitle={International Conference on Learning Representations},
  year={2022},
  doi={10.48550/arXiv.2106.09685}
}

@article{huang2024empirical,
  title={An empirical study of {LLaMA3} quantization: From {LLMs} to {MLLMs}},
  author={Huang, Wei and Zheng, Xingyu and Ma, Xudong and Qin, Haotong and Lv, Chengtao and Chen, Hong and Luo, Jie and Qi, Xiaojuan and Liu, Xianglong and Magno, Michele},
  journal={Visual Intelligence},
  volume={2},
  number={1},
  pages={Article 36},
  year={2024},
  doi={10.1007/s44267-024-00070-x}
}

@article{kane2013validating,
  title={Validating the interpretations and uses of test scores},
  author={Kane, Michael T.},
  journal={Journal of Educational Measurement},
  volume={50},
  number={1},
  pages={1--73},
  year={2013},
  doi={10.1111/jedm.12000}
}

@misc{kurt2026quantization,
  title={Which quantization should {I} use? {A} unified evaluation of llama.cpp quantization on {Llama}-3.1-8{B}-{Instruct} [Preprint]},
  author={Kurt, Uygar},
  howpublished={arXiv},
  year={2026},
  doi={10.48550/arXiv.2601.14277}
}

@article{latif2024finetuning,
  title={Fine-tuning {ChatGPT} for automatic scoring},
  author={Latif, Ehsan and Zhai, Xiaoming},
  journal={Computers and Education: Artificial Intelligence},
  volume={6},
  pages={Article 100210},
  year={2024},
  doi={10.1016/j.caeai.2024.100210}
}

@inproceedings{liew2024automated,
  title={On automated essay grading using large language models},
  author={Liew, Pei Yee and Tan, Ian K. T.},
  booktitle={Proceedings of the 2024 8th International Conference on Computer Science and Artificial Intelligence},
  pages={204--211},
  year={2024},
  publisher={Association for Computing Machinery},
  doi={10.1145/3709026.3709030}
}

@article{liu2025comparing,
  title={Comparing {GPT}-based approaches in automated writing evaluation},
  author={Liu, Yingying and Lu, Xiaofei and Qi, Huilei},
  journal={Assessing Writing},
  volume={66},
  pages={Article 100961},
  year={2025},
  doi={10.1016/j.asw.2025.100961}
}

@article{liu2025enhancing,
  title={Enhancing {GPT}-based automated essay scoring: The impact of fine-tuning and linguistic complexity measures},
  author={Liu, Yingying and Qi, Huilei and Lu, Xiaofei},
  journal={Computer Assisted Language Learning},
  year={2025},
  note={Advance online publication},
  doi={10.1080/09588221.2025.2518430}
}

@article{mizumoto2023exploring,
  title={Exploring the potential of using an {AI} language model for automated essay scoring},
  author={Mizumoto, Atsushi and Eguchi, Masaki},
  journal={Research Methods in Applied Linguistics},
  volume={2},
  number={2},
  pages={Article 100050},
  year={2023},
  doi={10.1016/j.rmal.2023.100050}
}

@misc{ormerod2024automated,
  title={Automated text scoring in the age of generative {AI} for the {GPU}-poor [Preprint]},
  author={Ormerod, Christopher Michael and Kwako, Alexander},
  howpublished={arXiv},
  year={2024},
  doi={10.48550/arXiv.2407.01873}
}

@article{ramesh2022automated,
  title={An automated essay scoring systems: A systematic literature review},
  author={Ramesh, Dadi and Sanampudi, Suresh Kumar},
  journal={Artificial Intelligence Review},
  volume={55},
  number={3},
  pages={2495--2527},
  year={2022},
  doi={10.1007/s10462-021-10068-2}
}

@article{ramineni2013automated,
  title={Automated essay scoring: Psychometric guidelines and practices},
  author={Ramineni, Chaitanya and Williamson, David M.},
  journal={Assessing Writing},
  volume={18},
  number={1},
  pages={25--39},
  year={2013},
  doi={10.1016/j.asw.2012.10.004}
}

@inproceedings{rathore2025rank,
  title={How much is too much? {Exploring} {LoRA} rank trade-offs for retaining knowledge and domain robustness},
  author={Rathore, Darshita and Kumar, Vineet and Bansal, Chetna and Moitra, Anindya},
  booktitle={Findings of the Association for Computational Linguistics: IJCNLP-AACL 2025},
  pages={1003--1013},
  year={2025},
  publisher={Association for Computational Linguistics},
  doi={10.18653/v1/2025.findings-ijcnlp.58}
}

@article{tang2024harnessing,
  title={Harnessing {LLMs} for multi-dimensional writing assessment: Reliability and alignment with human judgments},
  author={Tang, Xiaoyi and Chen, Hongwei and Lin, Daoyu and Li, Kexin},
  journal={Heliyon},
  volume={10},
  number={14},
  pages={Article e34262},
  year={2024},
  doi={10.1016/j.heliyon.2024.e34262}
}

@article{wang2024effectiveness,
  title={Effectiveness of large language models in automated evaluation of argumentative essays: Finetuning vs. zero-shot prompting},
  author={Wang, Qiao and Gayed, John Maurice},
  journal={Computer Assisted Language Learning},
  volume={39},
  number={1--2},
  pages={209--237},
  year={2024},
  doi={10.1080/09588221.2024.2371395}
}

@article{wang2014monitoring,
  title={Monitoring of scoring using the e-rater{\textregistered} automated scoring system and human raters on a writing test},
  author={Wang, Zhen and von Davier, Alina A.},
  journal={ETS Research Report Series},
  volume={2014},
  number={1},
  pages={1--21},
  year={2014},
  doi={10.1002/ets2.12005}
}

@article{williamson2012framework,
  title={A framework for evaluation and use of automated scoring},
  author={Williamson, David M. and Xi, Xiaoming and Breyer, F. Jay},
  journal={Educational Measurement: Issues and Practice},
  volume={31},
  number={1},
  pages={2--13},
  year={2012},
  doi={10.1111/j.1745-3992.2011.00223.x}
}

@article{xi2010automated,
  title={Automated scoring and feedback systems: Where are we and where are we heading?},
  author={Xi, Xiaoming},
  journal={Language Testing},
  volume={27},
  number={3},
  pages={291--300},
  year={2010},
  doi={10.1177/0265532210364643}
}

@inproceedings{zheng2024llamafactory,
  title={{LlamaFactory}: Unified efficient fine-tuning of 100+ language models},
  author={Zheng, Yaowei and Zhang, Richong and Zhang, Junhao and Ye, Yanhan and Luo, Zheyan},
  booktitle={Proceedings of the 62nd Annual Meeting of the Association for Computational Linguistics (Volume 3: System Demonstrations)},
  pages={400--410},
  year={2024},
  publisher={Association for Computational Linguistics},
  doi={10.18653/v1/2024.acl-demos.38}
}

\appendix
\titleformat{\section}{\normalfont\Large\bfseries}{Appendix~\thesection}{1em}{}

\section{System Prompts}
\label{app:prompts}

This appendix reproduces the two system prompts used in this study. The fine-tuning prompt (Appendix~\ref{app:prompts:training}) is the one shown to the model during LoRA fine-tuning, where the assistant's target output was a single numeric score. The production prompt (Appendix~\ref{app:prompts:production}) is the one used by the deployed AiAWE platform, which requests both a score and structured qualitative feedback across six dimensions. The fine-tuned model handles the production prompt without any additional training on feedback examples, evidencing the dual-purpose design described in Section~\ref{sec:finetuning}: the LoRA adapter contributes the scoring behavior, while the preserved instruction-following capabilities of the base model handle the richer feedback formatting at inference time.

\subsection{Fine-tuning system prompt (score-only target)}
\label{app:prompts:training}

\begin{tcolorbox}[colback=gray!5,colframe=gray!50!black,title=Fine-tuning system prompt,fonttitle=\bfseries,breakable]
\small
\begin{verbatim}
As a language expert, your task is to evaluate argumentative essays
on a scale of 0 to 5 (with 0.5 increments) based on the rubrics
below.

5 points:
- Effectively addresses the topic and task
- Is well organized and well developed, using clearly appropriate
  explanations, exemplifications and/or details
- Displays unity, progression and coherence
- Displays consistent facility in the use of language, demonstrating
  syntactic variety, appropriate word choice and idiomaticity, though
  it may have minor lexical or grammatical errors

4 points:
- Addresses the topic and task well, though some points may not be
  fully elaborated
- Is generally well organized and well developed, using appropriate
  and sufficient explanations, exemplifications and/or details
- Displays unity, progression and coherence, though it may contain
  occasional redundancy, digression, or unclear connections
- Displays facility in the use of language, demonstrating syntactic
  variety and range of vocabulary, though it will probably have
  occasional noticeable minor errors in structure, word form or use
  of idiomatic language that do not interfere with meaning

3 points:
- Addresses the topic and task using somewhat developed explanations,
  exemplifications and/or details
- Displays unity, progression and coherence, though connection of
  ideas may be occasionally obscured
- May demonstrate inconsistent facility in sentence formation and
  word choice that may result in lack of clarity and occasionally
  obscure meaning
- May display accurate but limited range of syntactic structures and
  vocabulary

2 points:
- Limited development in response to the topic and task
- Inadequate organization or connection of ideas
- Inappropriate or insufficient exemplifications, explanations or
  details to support or illustrate generalizations in response to
  the task
- A noticeably inappropriate choice of words or word forms
- An accumulation of errors in sentence structure and/or usage

1 point:
- Serious disorganization or underdevelopment
- Little or no detail, or irrelevant specifics, or questionable
  responsiveness to the task
- Serious and frequent errors in sentence structure or usage

0 points:
- An essay at this level merely copies words from the topic, rejects
  the topic, or is otherwise not connected to the topic, is written
  in a foreign language, consists of keystroke characters, or is
  blank.
\end{verbatim}
\end{tcolorbox}

During fine-tuning, the assistant's target output for each training example was simply a numeric score (e.g., \texttt{3.5}). No reasoning field, JSON formatting, or qualitative feedback was supplied as a target.

\subsection{Production system prompt (score + feedback)}
\label{app:prompts:production}

\begin{tcolorbox}[colback=gray!5,colframe=gray!50!black,title=Production system prompt,fonttitle=\bfseries,breakable]
\small
\begin{verbatim}
As a language expert, your task is to evaluate argumentative essays
on a scale of 0 to 5 (with 0.5 increments), based on the TOEFL
Independent Writing Task rubric. Please consider the essay topic
and the response given by the user carefully. The essay must
address the essay prompt in order to receive a score above one.
As a reference, the rubric is as below:

[Identical 0-5 rubric descriptors as in Appendix A.1]

Please return your evaluation as { "score": ..., } and detailed
feedback highlighting specific examples from the essay that need
improvement as { "reasoning": ... } in JSON format.

The feedback you enter in { "reasoning": ... } should include
recommended corrections on
  1) Surface errors,
  2) Formality and word choice,
  3) Sentence structure,
  4) Task completion,
  5) Coherence and cohesion, and lastly
  6) Overall justification for the score you gave.

Insert a "\n\n" in the reasoning you output to separate each point.

Please do not include markdown formatting in your response.
\end{verbatim}
\end{tcolorbox}

The production prompt instructs the model to produce JSON output with two fields: \texttt{score} (the numeric rating) and \texttt{reasoning} (a multi-paragraph string with structured feedback). The fine-tuned LoRA adapter has never been trained against this output format, but the model returns well-formed output in this format because LoRA leaves the base weights frozen and adds only a low-rank update, which largely preserves the base model's general instruction-following. Whether the adapter is kept as a separate file or merged into the base weights does not affect this, since the two give the same forward pass.

\section{Training Data Structure}
\label{app:training_data}

This appendix illustrates the structure of fine-tuning instances used to adapt the base models. The actual essays come from the Educational Testing Service (ETS) TOEFL Independent Writing dataset and cannot be reproduced here for licensing reasons. Instead, we provide synthetic examples that show the format precisely while making clear that the content is fabricated.

Each training instance is a JSON object with three messages (\texttt{system}, \texttt{user}, and \texttt{assistant}) following the standard chat-format conventions used by the LLaMA-Factory framework and most contemporary instruction-tuned LLMs. The \texttt{system} message contains the fine-tuning rubric prompt (Appendix~\ref{app:prompts:training}). The \texttt{user} message contains the essay prompt and the student response. The \texttt{assistant} message contains the human-assigned score from the ETS dataset, expressed as a number with one decimal place.

\subsection{Example 1: low-scoring synthetic essay}

\begin{tcolorbox}[colback=blue!2,colframe=blue!40!black,title={Synthetic training instance (illustrative, not real ETS data)},fonttitle=\bfseries,breakable]
\small
\begin{verbatim}
[
  {
    "role": "system",
    "content": "[Fine-tuning system prompt — see Appendix A.1]"
  },
  {
    "role": "user",
    "content": "Essay prompt: `Do you agree or disagree with the
following statement? Reading books made of paper is better than
reading books on electronic devices. Use specific reasons and
examples to support your answer.`
Essay content: `I agree to that question. Paper book is better.
Because eyes is tired from screen and battery dies sometime.
Also book smell nice and feels good. My grandmother she give me
old book and i like it very much. Therefore paper book win against
electronic.`"
  },
  {
    "role": "assistant",
    "content": "1.5"
  }
]
\end{verbatim}
\end{tcolorbox}

\subsection{Example 2: mid-range synthetic essay}

\begin{tcolorbox}[colback=blue!2,colframe=blue!40!black,title={Synthetic training instance (illustrative, not real ETS data)},fonttitle=\bfseries,breakable]
\small
\begin{verbatim}
[
  {
    "role": "system",
    "content": "[Fine-tuning system prompt — see Appendix A.1]"
  },
  {
    "role": "user",
    "content": "Essay prompt: `Do you agree or disagree with the
following statement? Working in a team is more productive than
working alone. Use specific reasons and examples to support your
answer.`
Essay content: `I agree that working in a team is generally more
productive than working alone, although there are some situations
where individual work is better. Teams allow members to share ideas
and divide tasks, which makes complex projects easier to complete.
For example, in my university group assignment last semester, we
finished a research report in two weeks because each member focused
on a different section.

However, teamwork also has challenges, such as scheduling conflicts
and disagreements about direction. When team members do not
communicate well, the work can become slower than if a single person
had done it. Therefore, while teams can be very effective, they
require good organization and clear communication to reach their
full productivity.`"
  },
  {
    "role": "assistant",
    "content": "3.5"
  }
]
\end{verbatim}
\end{tcolorbox}

\subsection{Notes on the training format}

Several details of this format are worth noting for researchers wishing to replicate the procedure:

\begin{itemize}
\item \textbf{The assistant's target is the score only.} No reasoning, JSON wrapper, or feedback text appears in the assistant message during fine-tuning. The model learns to map (rubric + essay prompt + essay) directly to a numeric score. The richer JSON output produced by the deployed system in response to the production prompt is generated by the base model's preserved instruction-following at inference time, not learned during fine-tuning.

\item \textbf{Score formatting.} Scores were stored as plain decimal strings (e.g., \texttt{2.0}, \texttt{3.5}, \texttt{4.5}). We did not pad, quantize, or class-encode the scores. The model is trained as a free-form text generator that happens to produce a number.

\item \textbf{Essay prompts are included in the user message.} Although the underlying ETS dataset uses only two essay prompts across all 480 essays, we include the prompt in every training instance rather than relying on the model to memorize it. 

\item \textbf{No data augmentation.} We did not paraphrase, back-translate, or otherwise augment the training data. The 120-essay training subset was used exactly as released by ETS.
\end{itemize}

The synthetic essays above are shorter than typical TOEFL Independent Writing responses (which average around 214 words in our dataset; see Section~\ref{sec:dataset}) and have been written specifically for this appendix to illustrate the format. Researchers seeking to fine-tune similar systems on their own essay datasets should adopt this format with their own prompts, essays, and scores.
\section{LoRA Mechanism}
\label{app:lora_fig}

\begin{figure}[!h]
\centering
\includegraphics[width=0.95\textwidth]{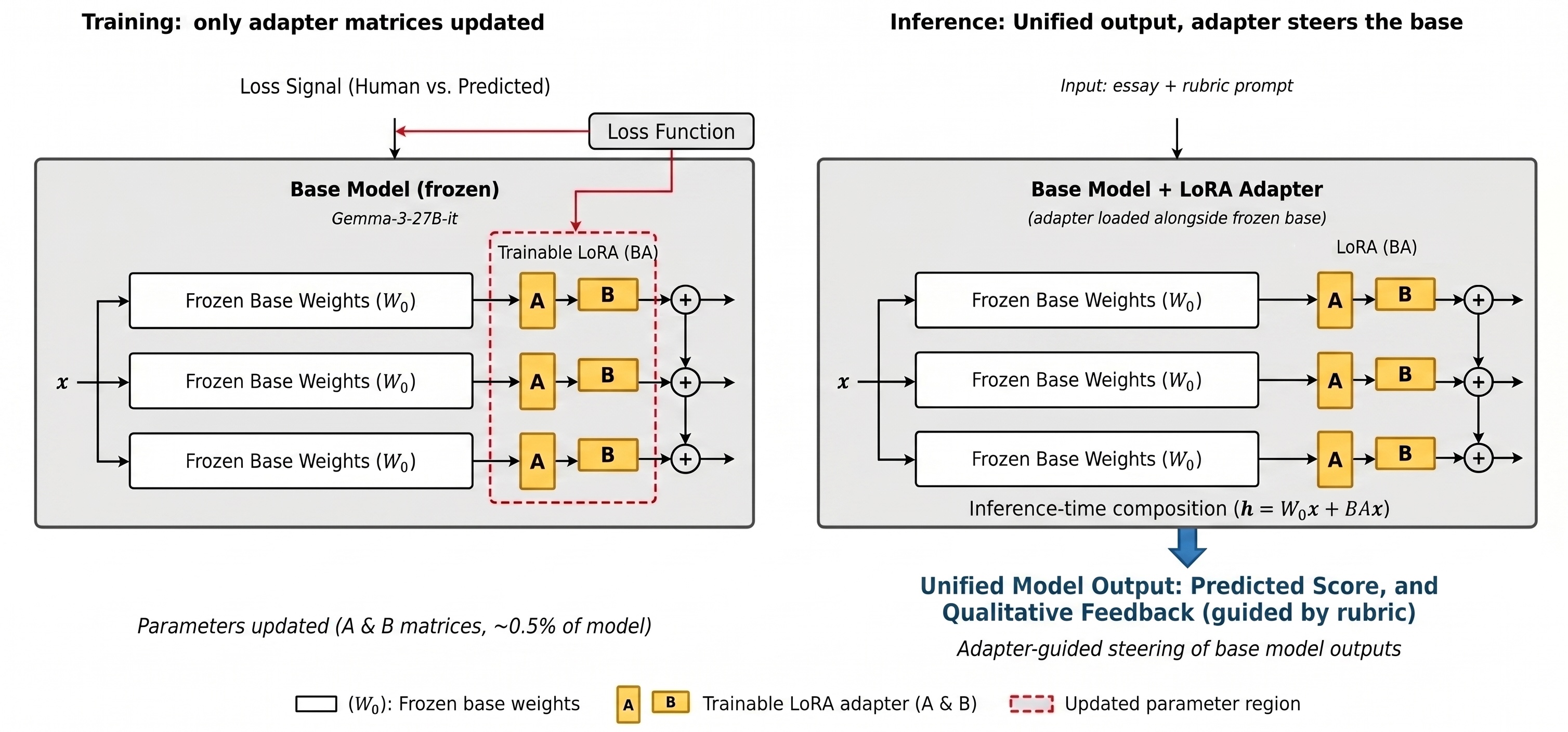}
\caption{A simplified explanation of LoRA. \textbf{Left:} during training, the base model's weights remain frozen. Gradients flow only into a small pair of trainable adapter matrices inserted into selected layers. \textbf{Right:} at inference time, the frozen base and the trained adapter produce the task-specialized output together.}
\label{fig:lora_explainer}
\end{figure}

\section{Quantization Details and Provenance}
\label{app:quant}

\subsection{Per-tensor precision allocation}

Table~\ref{tab:quant_tensors} shows the mixed-precision allocation of the \texttt{Q4\_K\_M} scheme in practice, based on inspection of the \texttt{bartowski} Gemma-3-27B-it GGUF file used in this study.

\begin{table}[!h]
\centering
\caption{Per-tensor precision allocation in \texttt{Q4\_K\_M} vs.\ \texttt{Q6\_K} GGUF quantization, based on inspection of the \texttt{bartowski} Gemma-3-27B-it files. Value and feed-forward down-projections are Q6\_K in 30 of the 62 transformer blocks (blocks 0--6, every third block from 9 to 54, and 55--61) and Q4\_K in the remaining 32. All other tensor types are the same in every block. Gemma 3 uses the token embedding matrix as its output layer, so there is no separate output tensor. Effective precision is approximately 4.9 bits per weight for Q4\_K\_M and 6.6 for Q6\_K. Norm layers remain at full precision in both schemes.}
\label{tab:quant_tensors}
\begin{tabular}{lcc}
\toprule
\textbf{Tensor type} & \textbf{Q4\_K\_M} & \textbf{Q6\_K} \\
\midrule
Token embeddings (\texttt{token\_embd}) & Q6\_K & Q6\_K \\
\midrule
Query projection (\texttt{attn\_q}) & Q4\_K & Q6\_K \\
Key projection (\texttt{attn\_k}) & Q4\_K & Q6\_K \\
Value projection (\texttt{attn\_v}) & Q6\_K\,/\,Q4\_K & Q6\_K \\
Output projection (\texttt{attn\_output}) & Q4\_K & Q6\_K \\
\midrule
FFN gate (\texttt{ffn\_gate}) & Q4\_K & Q6\_K \\
FFN up-projection (\texttt{ffn\_up}) & Q4\_K & Q6\_K \\
FFN down-projection (\texttt{ffn\_down}) & Q6\_K\,/\,Q4\_K & Q6\_K \\
\midrule
All norm layers & F32 & F32 \\
\midrule
\textbf{Approx.\ model size (Gemma-3-27B)} & \textbf{$\sim$16.5 GB} & \textbf{$\sim$22 GB} \\
\bottomrule
\end{tabular}
\end{table}

\subsection{Provenance}

The allocation in Table~\ref{tab:quant_tensors} is not a fixed property of the ``Q4\_K\_M'' label. It is set by the mixture rules of the specific quantization tool and version, which determine which tensors are promoted to higher precision, together with any explicit per-tensor overrides. The \texttt{bartowski} repositories generate GGUF files using \texttt{llama.cpp}'s quantization routines with an \emph{importance matrix} (imatrix), a calibration step in which a representative text dataset is passed through the unquantized model to measure which weights contribute most to output quality. The importance matrix does not change which tensors are promoted. It changes how the weights within each tensor are rounded. Two Q4\_K\_M files of the same base model, produced by different contributors, may therefore differ in their tensor allocation (if made with different tool versions or overrides), in their rounding (if made with different calibration data), and in how well quality is preserved on a given downstream task. Researchers replicating this study should use the exact \texttt{bartowski} artifacts we specify, or verify that their chosen quantization produces comparable scoring accuracy before drawing conclusions about the pipeline.

\subsection{Considerations for practitioners choosing a scheme}

The GGUF format supports more than a dozen schemes from roughly 1.5 to 8 bits per weight, and separate method families (GPTQ, AWQ, EXL2, bitsandbytes NF4) have their own trade-offs \citep{huang2024empirical, kurt2026quantization}. We selected Q4\_K\_M because it was among the most widely used 4-bit GGUF schemes available for our base models and suited our memory constraints. The available options change quickly, and our artifacts may not remain the best choice. Practitioners should evaluate against their own constraints, considering: available VRAM (the model, adapter, KV cache and working buffers must all fit); task sensitivity (fine-grained numerical reasoning and long-context coherence degrade faster than short-output scoring); the quality--speed trade-off (interactive use favors speed, batch processing favors quality); availability of well-calibrated community checkpoints; breadth of inference-framework support; and architecture-dependent behavior, since two models at the same nominal bit-width can lose very different amounts of quality. In short, quantization is an engineering choice to be tested empirically on the target task and hardware, not inherited from another study.

\section{Platform Implementation}
\label{app:platform}

AiAWE is deployed as a Django web application at \url{https://app.awade.gec.waseda.ac.jp}. The following describes the educator-facing features supporting institutional deployment. These are engineering context rather than evaluated contributions.

\paragraph{User and course management.} Administrators create courses and assign students through the Django admin interface. A batch user-creation tool accepts an Excel file of student email addresses, names and course assignments, provisioning accounts and enrollments in one step.

\paragraph{Batch essay processing.} Educators can upload an Excel file with essay prompt and essay text columns. The system processes all entries against a selected model and returns a downloadable file augmented with model scores, the \texttt{reasoning} field, processing timestamps and any error messages. The interface supports all configured models with per-model daily request quotas.

\paragraph{Custom rubric configuration.} New model configurations are created through the admin panel by specifying a system prompt, a user prompt template with placeholders for essay topic and text, and a temperature setting (0 for the deployed TOEFL scorer, giving deterministic output). This is the mechanism used to deploy the AWADE experimental model (Section~\ref{sec:awade_experimental}), and it allows an instructor to define a rubric for their own course without modifying the model or code.

\paragraph{API key management and quotas.} The platform uses token-based authentication with per-model daily request quotas, allowing administrators to manage access across courses and configurations without exhausting the shared inference hardware.

\section{Confusion Matrices}
\label{app:confusion}

Tables~\ref{tab:cm_gemma}--\ref{tab:cm_gpt} give the full confusion matrices for all three systems on the half-point grid ($N = 360$). Rows are the ETS composite score, columns the model prediction. Dots denote zero. Matrices on the collapsed integer grid follow by summing adjacent categories under the rounding rule given in Section~\ref{sec:metrics}.

\begin{table}[!h]
\centering
\caption{Confusion matrix, Gemma-3-27B (LoRA). Rows: ETS composite. Columns: predicted.}
\label{tab:cm_gemma}
\small
\begin{tabular}{c|ccccccccc|c}
\toprule
 & 1.0 & 1.5 & 2.0 & 2.5 & 3.0 & 3.5 & 4.0 & 4.5 & 5.0 & \textbf{$n$} \\
\midrule
1.0 & . & 2 & . & . & . & . & . & . & . & \textbf{2} \\
1.5 & . & 1 & 1 & . & . & . & . & . & . & \textbf{2} \\
2.0 & . & 1 & 13 & 6 & 1 & . & . & . & . & \textbf{21} \\
2.5 & . & . & 4 & 26 & 10 & . & . & . & . & \textbf{40} \\
3.0 & . & . & 3 & 35 & 50 & 15 & . & . & . & \textbf{103} \\
3.5 & . & . & . & 4 & 28 & 27 & 8 & 1 & . & \textbf{68} \\
4.0 & . & . & . & 1 & 8 & 20 & 16 & 8 & . & \textbf{53} \\
4.5 & . & . & . & . & . & 10 & 21 & 13 & 2 & \textbf{46} \\
5.0 & . & . & . & . & 1 & . & 5 & 9 & 10 & \textbf{25} \\
\midrule
\textbf{Total} & 0 & 4 & 21 & 72 & 98 & 72 & 50 & 31 & 12 & \textbf{360} \\
\bottomrule
\end{tabular}
\end{table}

\begin{table}[!h]
\centering
\caption{Confusion matrix, LLaMA-3.3-70B (LoRA). Rows: ETS composite. Columns: predicted.}
\label{tab:cm_llama}
\small
\begin{tabular}{c|ccccccccc|c}
\toprule
 & 1.0 & 1.5 & 2.0 & 2.5 & 3.0 & 3.5 & 4.0 & 4.5 & 5.0 & \textbf{$n$} \\
\midrule
1.0 & . & . & 2 & . & . & . & . & . & . & \textbf{2} \\
1.5 & . & . & 2 & . & . & . & . & . & . & \textbf{2} \\
2.0 & . & . & 9 & 11 & 1 & . & . & . & . & \textbf{21} \\
2.5 & . & . & 9 & 16 & 14 & 1 & . & . & . & \textbf{40} \\
3.0 & . & . & 3 & 29 & 47 & 22 & 2 & . & . & \textbf{103} \\
3.5 & . & . & . & 6 & 17 & 31 & 14 & . & . & \textbf{68} \\
4.0 & . & . & . & 1 & 9 & 16 & 27 & . & . & \textbf{53} \\
4.5 & . & . & . & . & 1 & 9 & 32 & 4 & . & \textbf{46} \\
5.0 & . & . & . & . & . & 1 & 10 & 12 & 2 & \textbf{25} \\
\midrule
\textbf{Total} & 0 & 0 & 25 & 63 & 89 & 80 & 85 & 16 & 2 & \textbf{360} \\
\bottomrule
\end{tabular}
\end{table}

\begin{table}[!h]
\centering
\caption{Confusion matrix, fine-tuned GPT-3.5 \citep{wang2024effectiveness}. Rows: ETS composite. Columns: predicted.}
\label{tab:cm_gpt}
\small
\begin{tabular}{c|ccccccccc|c}
\toprule
 & 1.0 & 1.5 & 2.0 & 2.5 & 3.0 & 3.5 & 4.0 & 4.5 & 5.0 & \textbf{$n$} \\
\midrule
1.0 & . & . & 2 & . & . & . & . & . & . & \textbf{2} \\
1.5 & . & . & 1 & 1 & . & . & . & . & . & \textbf{2} \\
2.0 & . & . & 11 & 9 & . & 1 & . & . & . & \textbf{21} \\
2.5 & . & . & 10 & 25 & . & 5 & . & . & . & \textbf{40} \\
3.0 & . & . & 6 & 40 & 1 & 45 & 2 & 9 & . & \textbf{103} \\
3.5 & . & . & . & 6 & . & 38 & 9 & 15 & . & \textbf{68} \\
4.0 & . & . & . & 3 & 1 & 16 & 4 & 29 & . & \textbf{53} \\
4.5 & . & . & . & . & . & 3 & 5 & 28 & 10 & \textbf{46} \\
5.0 & . & . & . & . & . & . & 1 & 13 & 11 & \textbf{25} \\
\midrule
\textbf{Total} & 0 & 0 & 30 & 84 & 2 & 108 & 21 & 94 & 21 & \textbf{360} \\
\bottomrule
\end{tabular}
\end{table}

The matrices make visible a property that summary statistics obscure. The GPT-3.5 baseline effectively avoids two categories: it issues only 2 predictions at 3.0 and 21 at 4.0, while assigning 108 essays to 3.5 and 94 to 4.5. Its errors are consequently concentrated in systematic displacements to adjacent half-points rather than distributed around the diagonal. Both open-weight models produce smoother marginal distributions, with Gemma's the closest to the criterion marginals. This is the same phenomenon discussed in Section~\ref{sec:predrange}, and it illustrates why the confusion matrix is a more informative object than any single agreement coefficient computed from it.

\section{Latency Benchmarking Protocol}
\label{app:bench}

\subsection{Protocol}

The figures in Table~\ref{tab:latency} come from a dedicated measurement run rather than from sampled production traffic. All 360 held-out essays were submitted to the deployed scoring endpoint under the production system prompt (Appendix~\ref{app:prompts:production}), so that the measured workload is identical to the workload whose accuracy is reported in Section~\ref{sec:results}. The configuration was: \texttt{bartowski} \texttt{google\_gemma-3-27b-it-Q4\_K\_M.gguf}; \texttt{llama.cpp} server with a 4096-token context; a single RTX 3090 (the workstation holds two, but the $\sim$16 GB quantized model and its KV cache fit within one 24 GB card, and the second was not engaged); requests issued sequentially.

For each request the server reports prompt evaluation time and token count, generation time and token count, and total time. These yield token counts, prefill and generation throughput, and end-to-end latency directly, without instrumentation of the application layer. We report prefill time as time to first token: this is exact for non-streaming requests, in which no token can be emitted before prefill completes.

Two properties of the run are worth recording because they bear on how the numbers should be read. First, although the server was configured with two slots, only one was ever active, and no request began before its predecessor completed. The measurements therefore describe isolated requests with no queuing contribution. Second, the server's prompt cache was enabled, but no request achieved a cache hit: the minimum observed prefill cost was 0.88 ms per token against a median of 1.11, whereas a cache hit would be one to two orders of magnitude cheaper. Every reported prefill is a cold prefill, and the figures are not flattered by caching.

\subsection{What these measurements do not establish}

\paragraph{Concurrency scaling is not characterized.} Requests were issued sequentially, so we report throughput at concurrency 1 (4.0 essays per minute) but cannot report how throughput scales when multiple students submit simultaneously. On a two-slot server one would expect sub-linear gains as the two slots contend for memory bandwidth, but we have not measured this and do not assert it. The batch-processing figures an institution would need for capacity planning therefore remain unestablished, and a concurrency sweep is the obvious extension.

\paragraph{Single run, single hardware configuration.} These are measurements from one pass on one machine. We do not report run-to-run variance, and the figures should not be transferred to other GPUs, other quantization schemes, or other \texttt{llama.cpp} builds without re-measurement, for the reasons set out in Appendix~\ref{app:quant}.

\section{Resource Availability}
\label{app:availability}

The deployed AiAWE platform is publicly accessible at
\url{https://app.awade.gec.waseda.ac.jp}, providing essay scoring,
rubric-referenced feedback, and batch processing through a web interface.

The fine-tuned LoRA adapters used in this study are available on the Hugging Face
Hub under \url{https://huggingface.co/jgayed}, including the Gemma-3-27B-it
adapters trained on the 120-essay split \citep{jgayed_gemmalora120} and on the
full 480-essay set \citep{jgayed_gemmalorafull}, along with GGUF-quantized
variants and the corresponding LLaMA-3.3-70B adapters.

The application source code and the LLaMA-Factory training configuration (YAML)
files are released through the project's GitHub organization
(\url{https://github.com/waseda-awade}).

The 480-essay ETS TOEFL Independent Writing dataset cannot be redistributed for
licensing reasons. Researchers seeking access should contact Educational Testing
Service directly.
\end{document}